\documentclass[useAMS]{mn2e}
\usepackage{epsfig}

\title[The dust production rate of AGB stars in the MCs]{The dust production rate of AGB stars in the Magellanic Clouds}
\author[R. Schneider et al.]{Raffaella Schneider$^{1}$\thanks{E-mail:
raffaella.schneider@oa-roma.inaf.it}, Rosa Valiante$^{1}$, Paolo Ventura$^{1}$, Flavia dell'Agli$^1$, 
\newauthor
Marcella Di Criscienzo$^{1,2}$, Hiroyuki Hirashita$^3$ and Francisca Kemper$^3$\\
\\
$^{1}$INAF/Osservatorio Astronomico di Roma, Via di Frascati 33, 00040 Monteporzio, Italy\\
$^{2}$INAF/Osservatorio Astronomico di Capodimonte, Salita Moiarello 16, 80131 Napoli, Italy\\
$^{3}$Institute of Astronomy and Astrophysics, Academia Sinica, PO Box 23-141, Taipei 10617, Taiwan}

\begin{document}

\date{}

\pagerange{\pageref{firstpage}--\pageref{lastpage}} \pubyear{2014}

\maketitle

\label{firstpage}

\begin{abstract}
We compare theoretical dust yields for stars with mass $\rm 1\,M_{\odot} \le \rm m_{star} \le \rm 8 \, M_{\odot}$, and
metallicities $\rm 0.001 \le \rm Z \le \rm 0.008$ with observed dust production
rates (DPR) by carbon-rich and oxygen-rich Asymptotic Giant Branch (C-AGB and
O-AGB) stars
in the Large and Small Magellanic Clouds (LMC, SMC).
The measured DPR of C-AGB in the LMC are
reproduced only if the mass loss from AGB
stars is very efficient during the carbon-star stage. The
same yields
over-predict the observed DPR in the SMC,
suggesting a stronger metallicity dependence of the mass-loss rates
during the carbon-star stage.
DPR of O-AGB stars suggest that rapid silicate dust enrichment
occurs due to efficient hot-bottom-burning if $\rm m_{star} \ge \rm 3 \,
M_{\odot}$ and $\rm Z \ge 0.001$.
When compared to the most recent observations, our models support
a stellar origin for the existing dust mass, if no significant
destruction in the ISM occurs, with
a contribution from AGB stars of $70 \%$ in the LMC and $15 \%$ in the SMC.
\end{abstract}

\begin{keywords}
Stars: AGB and post-AGB, supernovae: general. ISM: dust, extinction. Galaxies: Magellanic Clouds, evolution, ISM 
\end{keywords}

\section{Introduction}

The best way to compare the total dust input from evolved stars in a galaxy to 
the total dust budget is to detect the entire population of dusty stars at infrared 
(IR) wavelengths and estimate the dust-injection rate of each. 
These global measurements are not possible in our Galaxy, due to source
confusion
in the Galactic plane, but have been attempted on nearby external galaxies, such as 
the Small and Large Magellanic Clouds (hereafter SMC and LMC, respectively). 
Extragalactic studies have also the advantage that distances to sources within a galaxy 
can be assumed to be the same.

The importance of the Magellanic Clouds as laboratories of dust enrichment by stellar sources
has been thoroughly discussed in the literature (Matsuura et al. 2009, 2013; Srinivasan et al. 2009; 
Boyer et al. 2012; Riebel et al. 2012). A growing body of 
observational data has been made available to the community by means of dedicated large photometric surveys.
Among the others, the Magellanic Clouds Photometric Survey (MCPS, Zaritsky et al. 2004), the Two Micron
All Sky Survey (2MASS, Skrutskie et al. 2006), Surveying the Agents of a Galaxy's Evolution
Survey (SAGE) with the {\it Spitzer} Space Telescope (SAGE-LMC, Meixner et al. 2006; SAGE-SMC, Gordon et al. 2011), and
{\it HERschel} Inventory of The Agents of Galaxy Evolution (HERITAGE, Meixner et al. 2010, 2013)
have provided catalogues of point sources as well as high-resolution maps of the emission by
the warm and cold dust components in the interstellar medium (ISM). 

In addition, a wealth of complementary data allows to recontruct the recent and past star formation
histories of the galaxies (see, among others, Harris \& Zaritsky 2004, 2009; Bolatto et al. 2011; Skibba et al. 2012;
Weisz et al. 2013; Cignoni et al. 2013), their metal enrichment histories (see e.g. Carrera et al. 2008a, 2008b; Piatti 2012;
Piatti \& Geisler 2013), and their present-day global gas, stellar and dust content (see Meixner et al.
2013 for a recent collection of observational data). 

Hence, these two galaxies represent an excellent astrophysical laboratory to investigate the life-cycle
of dust in the ISM, providing a fundamental benchmark to theoretical models.

Knowledge of the amount and composition of dust formed by stars of intermediate mass 
($\rm 1 M_{\odot} \le \rm m_{star} \le \rm 8 M_{\odot}$) and in the ejecta of core-collapse 
supernovae ($\rm m_{star} > 8 M_{\odot}$, where $\rm m_{star}$ is the stellar mass at zero-age main sequence) 
represents the first step towards the understanding
of dust enrichment from the most distant galaxies to the Local Universe. The relative importance
of these stellar sources of dust depends on the mass-dependent dust yields, on the stellar initial
mass function (IMF), as well as on the star formation history (SFH) of each galaxy (Valiante et al. 2009, 2011).   
Contrary to previous claims, dust production at high redshifts, $\rm z > 6$, is not necessarily 
driven by massive stars, as stars of intermediate mass on their Asymptotic Giant Branch (AGB)
can dominate dust production on a timescale which ranges between 150 and 500 Myr (Valiante et al. 2009).
Local observations of dust emission in the circumstellar shells of evolved stars or in 
recent supernovae and supernova remnants have the potential to significantly reduce uncertainties
associated to theoretical dust yields. 

In this paper, our main aim is to compare dust production rates by AGB stars 
calculated by means of theoretical models to observations of evolved stars in the Magellanic
Clouds. In particular, we consider a new grid of dust yields for different stellar masses and 
metallicities (Ventura et al. 2012a, 2012b, Di Criscienzo et al. 2013; Ventura et al. 2014)
which is based on models calculated with a full integration of the stellar structure, following the
evolution from the pre-main sequence phase using the code ATON (Ventura et al. 1998; Ventura \& D'Antona 
2009). This represents an important difference with respect to previous studies by Ferrarotti \& Gail, whose dust yields are based 
on stellar properties computed from synthetic\footnote{In this
context, by synthetic models we refer to models in which the evolution 
is described with analytical relations derived by fitting the results
of full evolutionary models.} models (Ferrarotti \&
Gail 2001, 2002, 2006; Zhukovska, Gail \& Trieloff 2008) or with respect to more recent hybrid models where the integration
is limited to the envelope structure of the stars (Marigo et al. 2013; Nanni et al. 2013). 
 As a consequence, the mass and metallicity dependence of carbon and 
silicate dust yields based on ATON stellar models can not be reproduced by dust yields based
on synthetic models, due to the different treatment of physical processes, such as the third dredge-up 
and the hot bottom burning, which alter the surface chemistry of AGB stars (Ventura et al. 2014). 

By comparing the predictions of different sets of AGB stars dust yields to observations of carbon-rich and oxygen-rich AGB stars in
the Magellanic Clouds, we can hope to constrain some of the model uncertainties. 
In addition, the difference in the gas metallicity of the two galaxies, with $\rm Z_{SMC} = 0.004$ and 
$\rm Z_{LMC} = 0.008$, allows to explore the complex and poorly understood dependence of AGB stars dust
production rates on their progenitors initial metallicity (Ventura et al. 2012b). Finally, we can assess the
relative importance of AGB stars and supernovae as dust producers in the two galaxies, comparing
the overall contribution of stellar sources to the estimated total dust mass in the ISM.

In a recent paper, Zhukovska \& Henning (2013) have done a similar analysis, discussing the dust
input from AGB stars in the LMC. For the first time, theoretically calculated dust production rates 
of AGB stars have been compared to those derived from IR observations of AGB stars for the entire galaxy. 
In their comparison, they consider synthetic yields by Zhukovska et al. (2008) but discuss also 
the implications of their models when ATON yields are adopted. They find that while synthetic models
lead to carbon and silicate dust production rates in good agreement with observations, 
ATON (hereafter {\bf old} ATON) models under-predict carbon-dust production rates, favouring silicate dust production, in contrast
to the observations. Motivated by these findings, we have recently investigated the dependence of
the predicted dust yields on the macrophysics adopted to describe the AGB evolution (Ventura et al. 2014)
and a new grid of ATON (hereafter {\bf new} ATON) dust yields has been 
computed for different initial metallicities, which range between $3 \times 10^{-4}$ to $8 \times 10^{-3}$, 
including the metallicity of the SMC, $\rm Z = 0.004$, 
that has not been considered in previous calculations.

Here we extend the comparison to this new ATON grid. In addition, we do not limit the analysis
to AGB stars in the LMC but we test the models against observations in the SMC.

The paper is organized as follows: in Section 2, we briefly summarize the AGB stellar dust yields predicted by different theoretical models;
in Section 3 we review  the observationally constrained 
star formation and chemical enrichment histories of the Magellanic Clouds; 
the associated dust production rates for different sets of dust yields are presented in Section 4 and compared to observational
data. The best fit models are then used, in Section 5, to assess the role of AGB stars and
supernovae in the global dust budget of the LMC and SMC. Finally, in Section 6 we summarize
and discuss our conclusions.  

\begin{figure*}
\includegraphics[width=58mm]{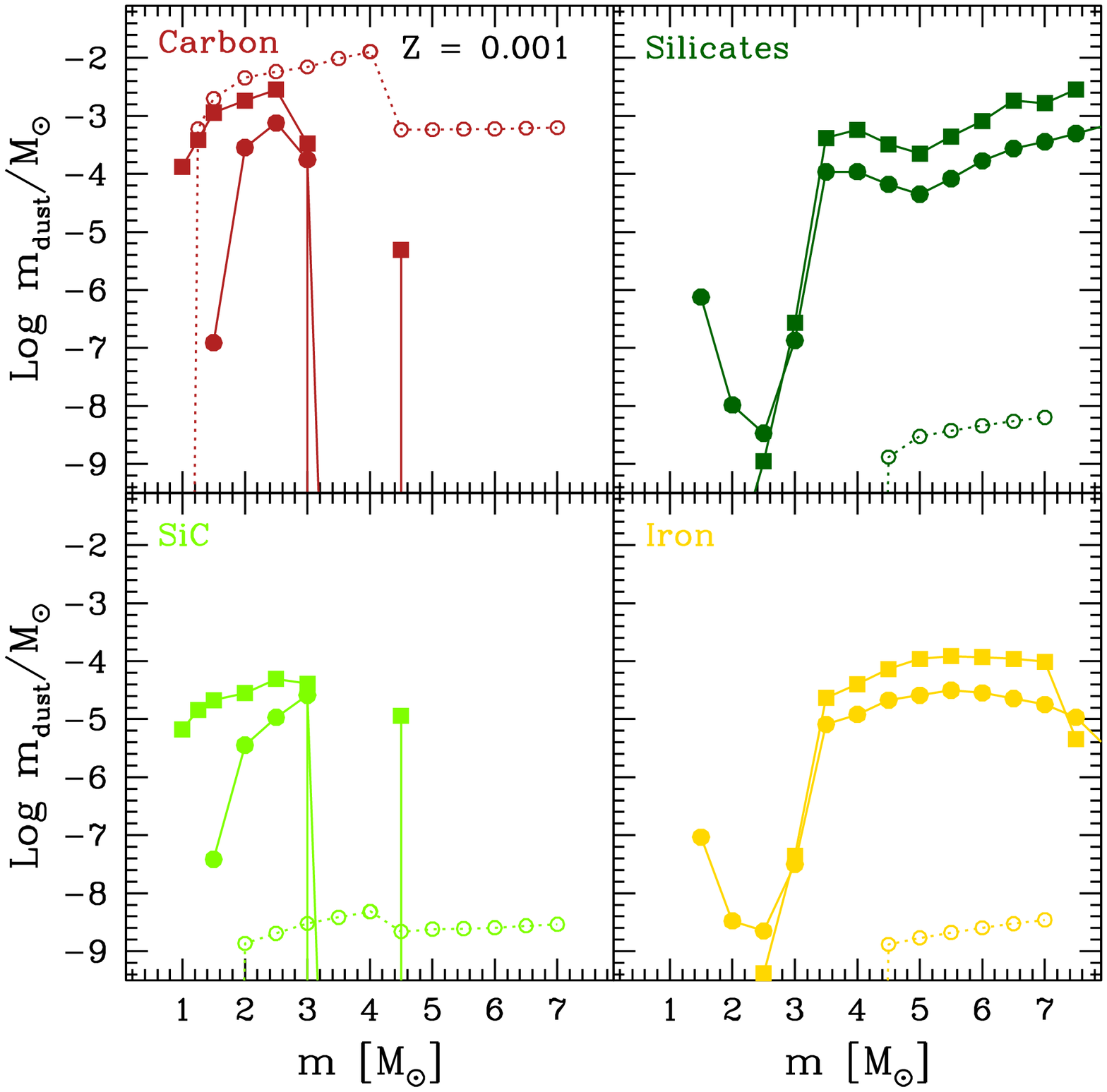}
\includegraphics[width=58mm]{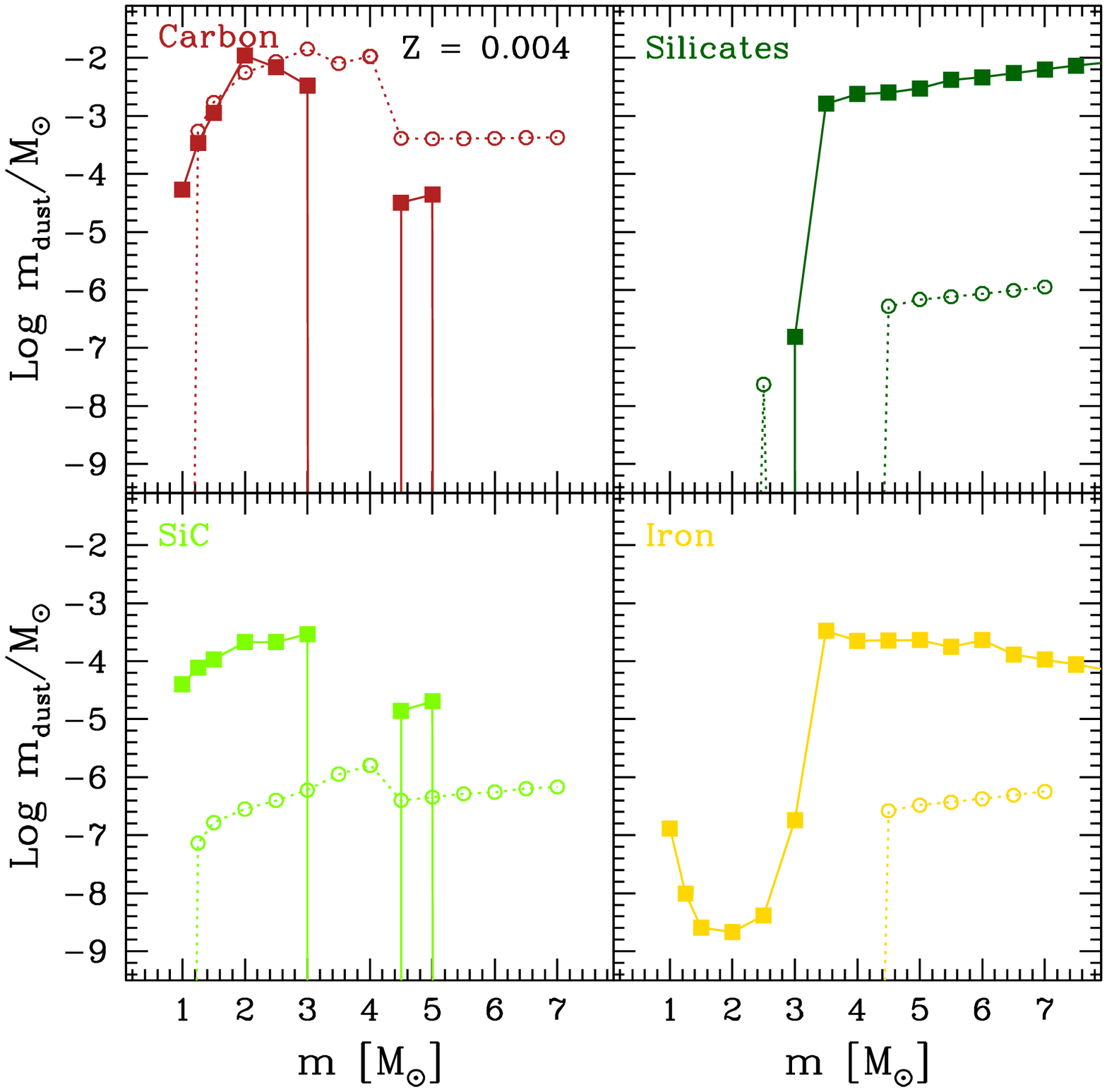}
\includegraphics[width=58mm]{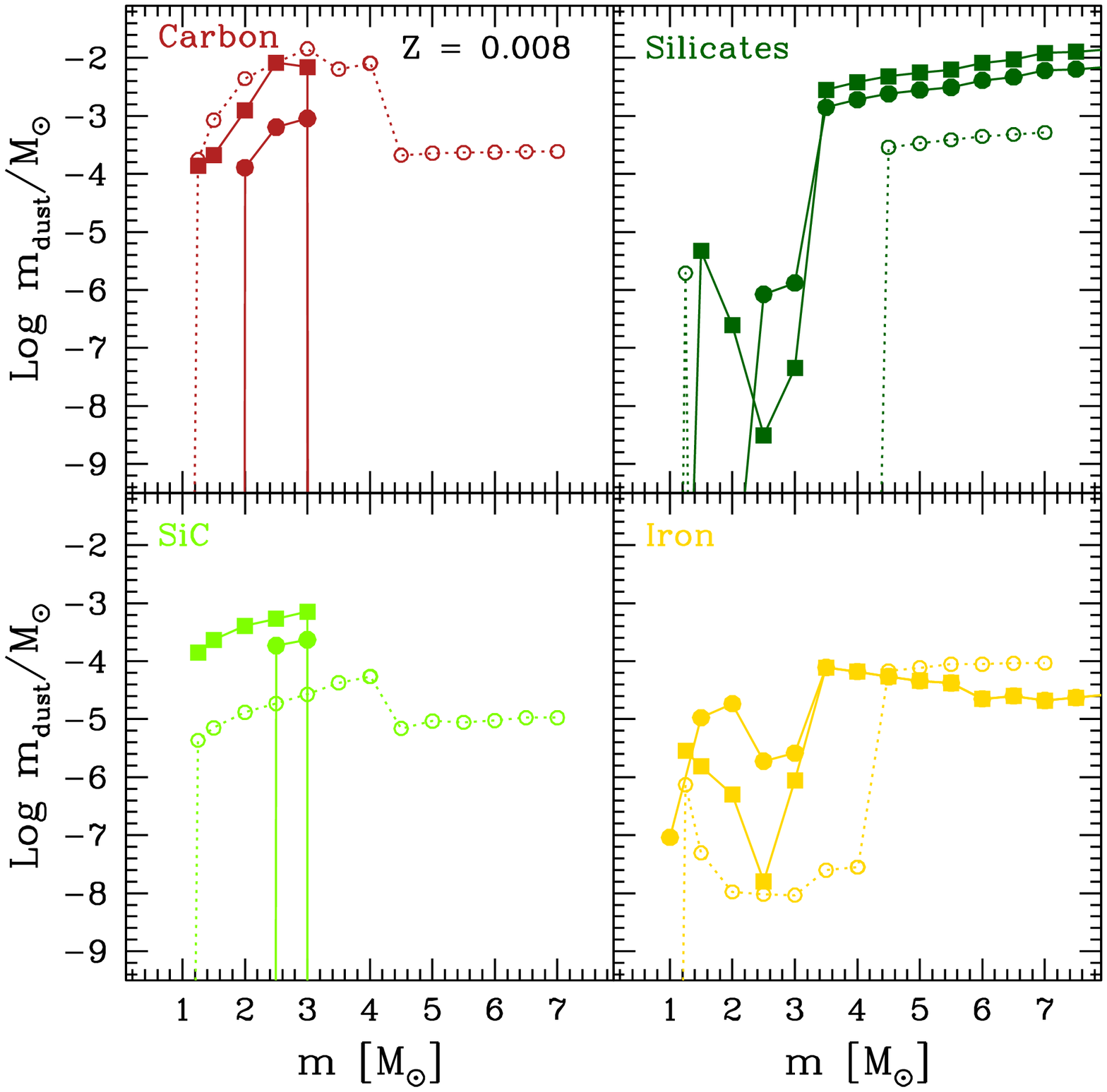}
\caption{Dust yields from AGB stars as a function of their initial mass for three initial metallicities: Z = 0.001, 0.004, and 0.008 (left, central, and right panels,
respectively), with separate contributions from carbon dust, silicates, SiC and Iron grains. In each panel, empty squares show the
predictions from Zhukovska et al. (2008), filled circles and squares are the old and new ATON yields, respectively (see text).}
\label{fig:agbyields}
\end{figure*}

\section{Dust yields from AGB stars}

Theoretical calculations of the total dust mass formed during the AGB phase has recently attracted many
independent studies.
In their pioneering work, Ferrarotti \& Gail (2001, 2002, 2006) used synthetic
stellar evolution models to compute the non-equilibrium grain condensation and estimated the dust
yields for M, S, and C-type AGB stars of different ages and metallicities 
(Zhukovska, Gail \& Trieloff 2008). These models have proved to be very useful tools, thanks to the
simplicity with which they can be incorporated into stellar population synthesis studies and chemical
evolution models (Zhukovska \& Henning 2013). However, their major drawback is to 
rely on simple evolutionary codes where some physical processes, such as  the variation of the core mass,
the temperature at the bottom of the convective envelope, the core mass at which the Third Dredge Up (TDU) begins to
operate, the extent of the inwards penetration of the surface convective zone, are described by means
of analytic approximations (Marigo \& Girardi 2007). Since these processes affect the chemical composition
and physical conditions in the circumstellar atmosphere during the AGB phase, they also affect the
resulting dust mass and composition.
 
To overcome this limitation, in a series of papers we have recently 
published a grid of AGB stars dust yields for different stellar masses and metallicities using models
that follow stellar evolution from the pre-main sequence phase until the almost complete ejection
of the stellar mantle by means of the code ATON (Ventura et al. 1998; Ventura \& D'Antona 2009). 
In the first paper of the series, Ventura et al. (2012a) computed the mass and composition of dust produced by stars with 
masses in the range $\rm 1 M_{\odot} \le m \le 8 M_{\odot}$ and with a metallicity of Z = 0.001 during 
their AGB and super-AGB phases. They found that the dust composition depends on the stellar mass:
low-mass stars, with $\rm m < 3 M_{\odot}$, produce carbon dust whereas more massive stars
experience Hot Bottom Burning, do not reach the C-star stage and produce silicates and iron grains.

Starting from a higher initial metallicity, Z = 0.008, Ventura et al. (2012b) found a similar
trend, with a transition between carbon and silicate dust production occurring at a threshold mass of 
$\rm 3.5 \, M_\odot$. While the yields of carbon dust do not depend on metallicity, the mass of silicate
dust grows with metallicity due to the combined effect of the softer HBB experienced 
and of the larger silicon abundance.

Conversely, for stars with initial metallicity Z $< 0.001$, no silicate dust formation occurs due to the scarcity
of silicon available in the envelope; carbon dust continue to form in stars with masses $\rm \le 2.5 M_{\odot}$ 
down to initial metallicities of Z $ = 3\times 10^{-4}$, below which even these low-mass stars experience 
HBB, do not become carbon stars and AGB stars do not produce any dust (Di Criscienzo et al. 2013).

It is important to stress that some of these results could not be captured by previous models as they heavily
depend on fundamental physical processes occurring during the AGB evolution. In particular, the TDU 
and the Hot Bottom Burning (HBB) alter the surface chemistry of AGB stars. Silicate dust production depends 
on the modelling of convection which, in turns, determines the strength of HBB. 
On the other hand, the mass of carbon dust formed depends on the extent of the third dredge-up: a small
amount of extra-mixing from the borders of the convective shell developed during the thermal pulses
favour a much larger penetration of the convective envelope, leading to a much stronger third dredge
up, hence a larger carbon dust production. These issues have been thoroughly discussed by Ventura et al. (2014),
who explored the impact of different physical assumptions concerning the extra-mixing and
mass-loss during the C-star phase on the resulting dust yields. 
\begin{figure*}
\includegraphics[width=80mm]{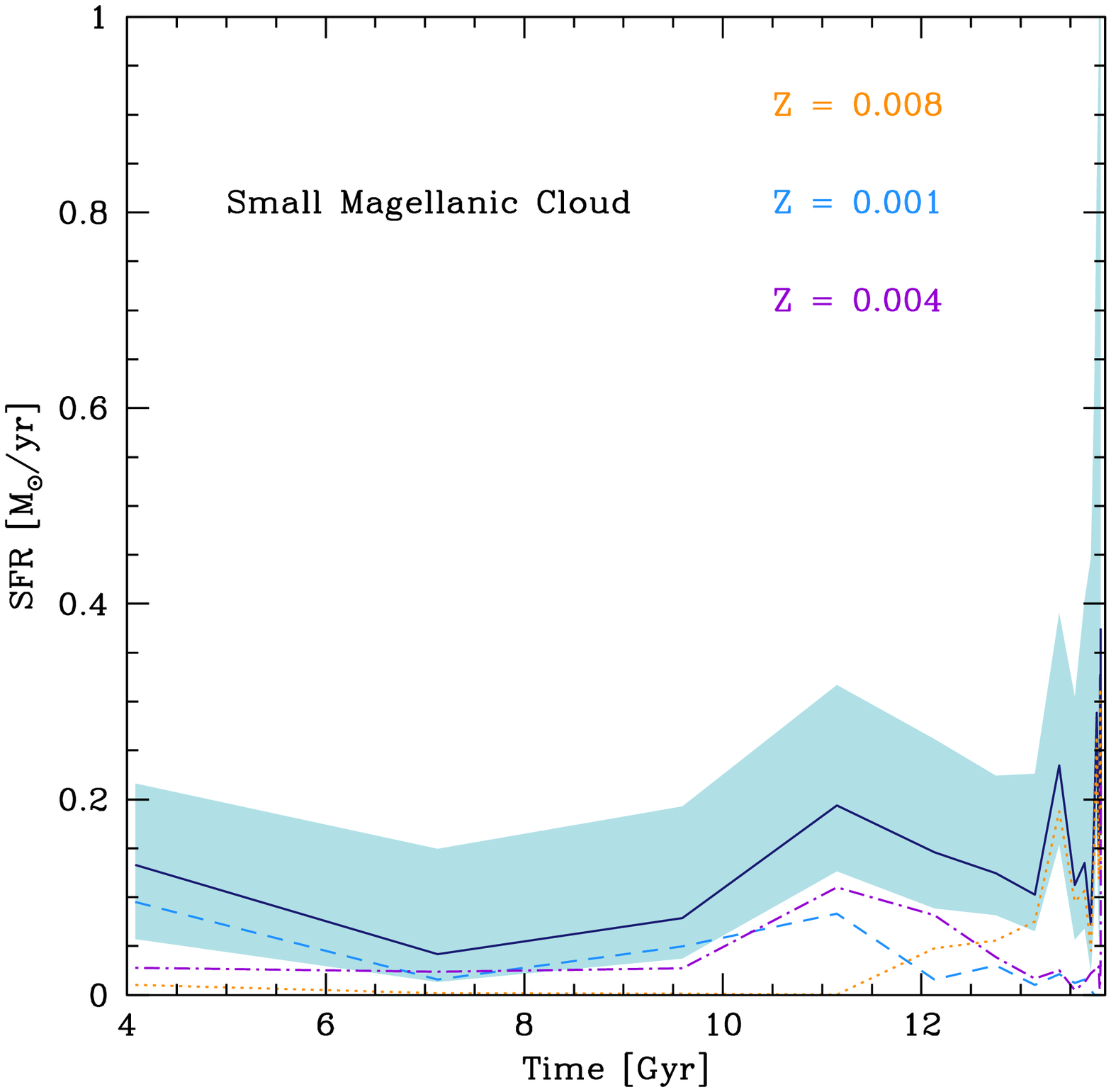}
\includegraphics[width=80mm]{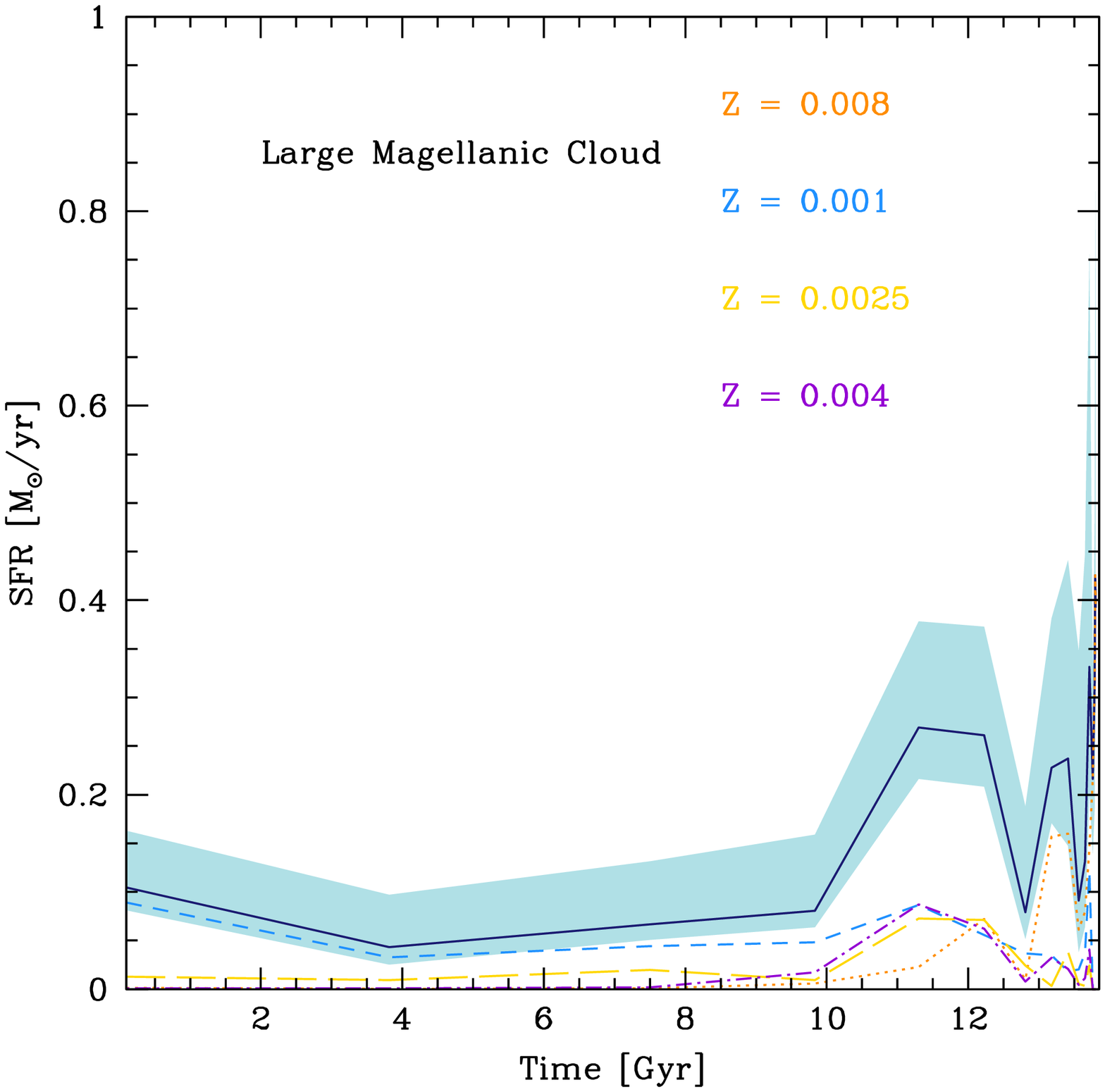}
\caption{The star formation history of the Small (left panel) and Large (right panel) Magellanic
Clouds as a function of time from Harris \& Zaritsky (2004, 2009). The solid lines show
the best-fit star formation history, with shaded regions representing the uncertainty on the fit. 
The separate contribution of different metallicity bins is also shown with Z = $0.001, 0.0025, 0.004,$
and $0.008$ plotted as dashed, long-dashed, dot-dashed, and dotted lines, respectively. The star formation
history for the SMC starts at 4 Gyr as the oldest stellar age bin in Harris \& Zaritsky (2004) is $\rm 9.988 Gyr$.}
\label{fig:SFH}
\end{figure*}
In Fig.~\ref{fig:agbyields}, we show the AGB stars dust yields predicted by the old and new ATON models. For
comparison, we also show the dust yields predicted by Zhukovska et al. (2008) using synthetic AGB stellar 
models (hereafter
Z08 AGB stars dust yields).
A detailed analysis of the differences between the different grids of dust yields has been given in
Ventura et al. (2014) and we refer the interested reader to the original paper for more details. Here we
limit the discussion to the features which are of interest for the purpose of our study, namely the 
mass and metallicity dependence of carbon dust and silicates. In particular, it is evident from Fig.~\ref{fig:agbyields}
that the new ATON models predict larger carbon (and SiC) dust yields with respect to the old ATON models, 
independent of the initial stellar metallicity. 
This is due to the effect of the extra-mixing (larger efficiency of TDU) and - to a larger extent - to the increased mass loss rates in the
C-star stage.  In the new ATON models mass loss during the C-star stage is described following the formulation
by Wachter et al. (2008) which accounts for carbon-dust production and the consequent acceleration of the wind. For
all the other evolutionary phases, the mass loss prescription by Bl{\"o}cker (1995) is used, similarly to the old
ATON models. The mass of silicates (and iron) dust produced by AGB stars with larger mass is larger in the new ATON
models, particularly for the lowest metallicity, due to the stronger HBB experienced.

The largest difference between ATON  and Z08 dust yields concerns the behaviour of AGB stars with initial masses 
$ > 3 \rm M_{\odot}$: these contribute to carbon-dust production according to Z08, but never reach the carbon-star stage 
according to ATON models. 
This difference is due to the different treatment of convection, that reflects into a much stronger HBB experienced by our models, 
in comparison to Z08, causing the difference in the predicted silicate dust production. It is important to note that
the mass segregation between carbon dust production by low-mass AGB stars and silicate dust production by 
high-mass AGB stars is a feature unique to our dust models, and it is not apparent in the models by Zhukovska et al. (2008)
nor in the most recent models by Nanni et al. (2013). We do not show in the figure the results of the 
latter study as these are qualitatively in good agreement with those of Zhukovska et al. (2008). In fact,
although Nanni et al. (2013) have based their calculation on a complete envelope model that represents 
a significant step forward with respect to purely synthetic AGB models, their approach does not
follow the complete stellar evolution. It is important to stress that the HBB phenomenon, which is responsible
for the large differences in the mass and composition of dust produced by high-mass AGB stars, is
a consequence of the delicate coupling between the outer region of the degenerate core, 
the CNO burning layer, and the innermost regions of the surface convective zone; {\it by definition, this
cannot be described within the framework of a synthetic modelling of the AGB phase.}

\section{The star formation and metal enrichment histories of the Magellanic Clouds}
\label{sec:sfr}

Spatially resolved star formation histories of the SMC and LMC have been
obtained by Harris \& Zaritsky (2004, 2009) based on the Magellanic
Clouds Photometric Survey (MCPS), which includes over 6 million SMC stars and
20 million LMC stars. The star formation history of each observed 
stellar population is obtained minimizing the difference between observed 
and synthetic color-magnitude diagrams (CMD) selected from a library generated
by means of the StarFISH code (Harris \& Zaritsky 2001) spanning appropriate
ranges in metallicity and ages. Adopting a Salpeter Initial Mass Function (IMF),
with a binary fraction of 0.5, the star formation rate as a function of the stellar
ages for 3 (4) metallicity bins has been inferred for 350 (1376) regions of the SMC (LMC).
In Fig.~\ref{fig:SFH} we show the total best-fit star formation history (solid lines, with the
shaded region representing the uncertainty on the fit) and the separate contribution of different
metallicity bins.    

\begin{figure*}
\includegraphics[width=58mm]{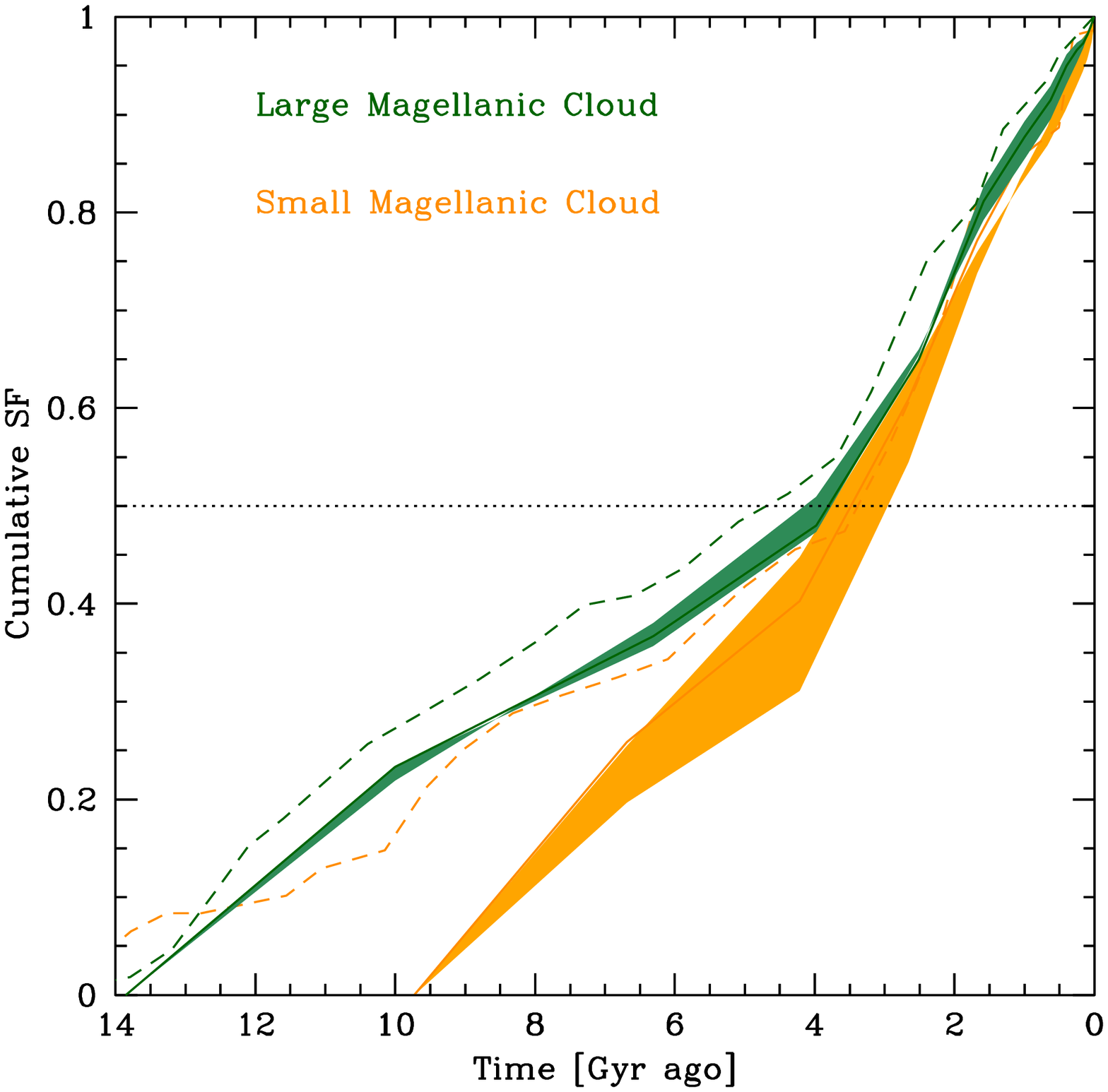}
\includegraphics[width=58mm]{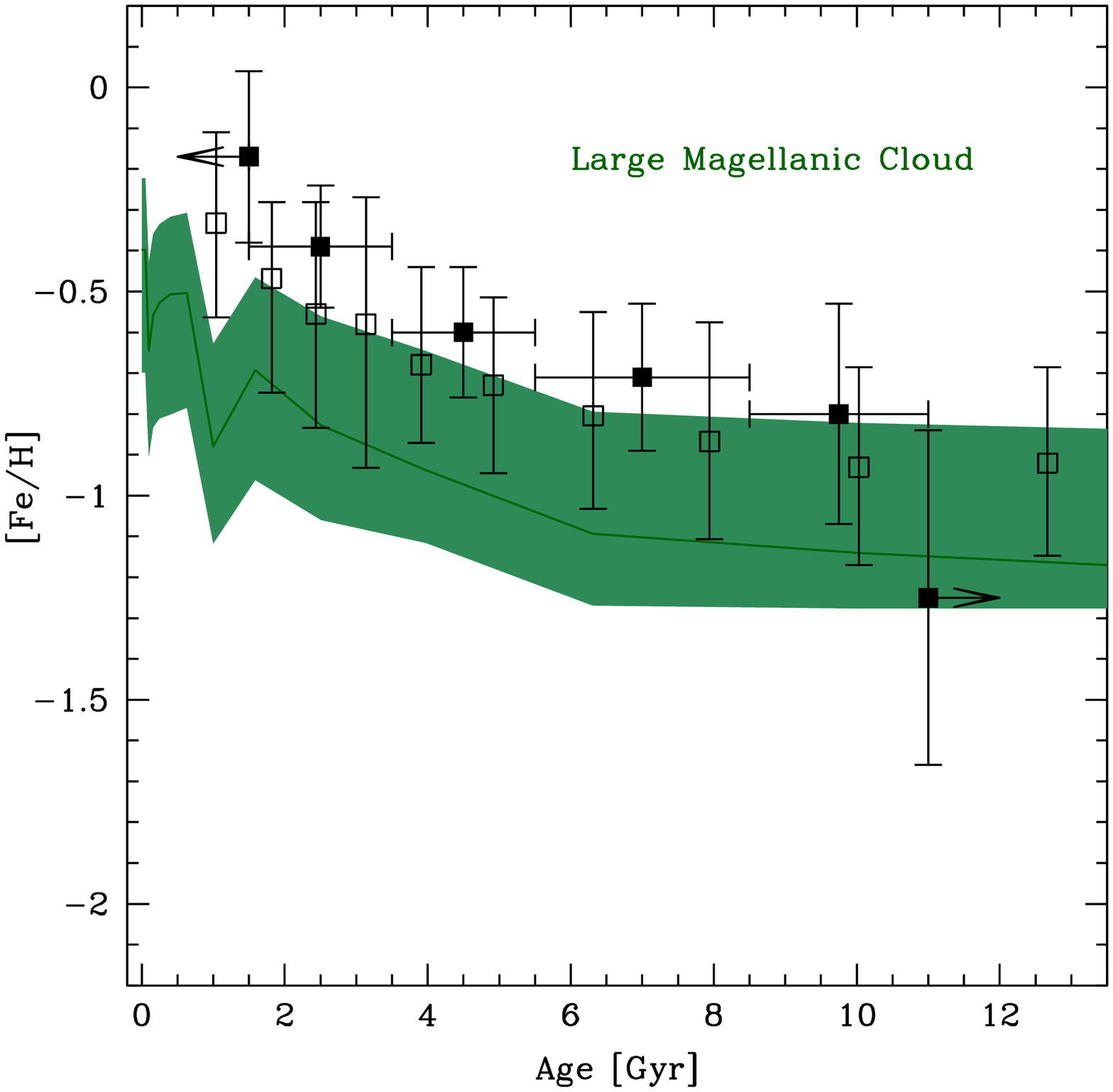}
\includegraphics[width=58mm]{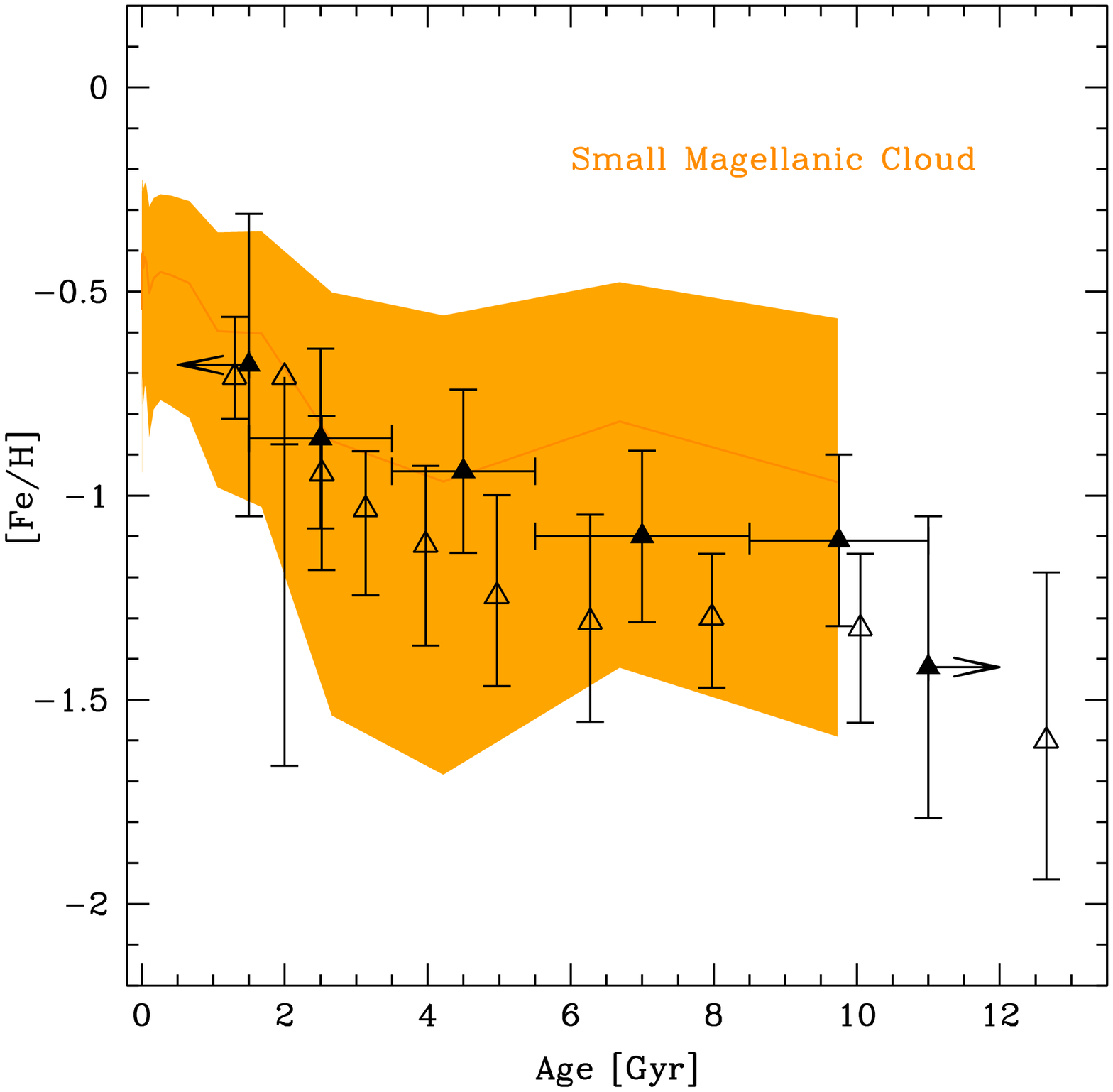}
\caption{The left panel shows the cumulative star formation histories of the Magellanic Clouds 
corresponding to the results shown in Fig.\ref{fig:SFH}. The dashed lines illustrate the recent
results of Weisz et al. (2013) using deep HST data (see text). For illustrative purposes, 
the horizontal dashed line indicates 50\% of the total stellar mass. The central and right panels
show the age-metallicity relation obtained from the best-fit solution of Harris \& Zaritsky (2004,
2009) for the Magellanic Clouds. Filled squares (triangles) are the average results obtained by
Carrera et al (2000a) for the LMC disk population (SMC); the average
results obtained by Piatti \& Geisler (2013) for the LMC and by Piatti (2012) for the SMC are shown
with empty squares and triangles, respectively.}
\label{fig:MCevo}
\end{figure*}

There are several temporally coincident features in the global star formation histories of the two
galaxies, suggesting a common evolution for the Magellanic Clouds. 
Both galaxies experienced a long ($\sim 5-7$~Gyr) quiescent epoch starting roughly 10 Gyr ago,
followed by a peak in the star formation rate roughly 2-3 Gyr ago and an enhanced star
formation activity around 400 Myr ago (Harris \& Zaritsky 2004, 2009). The best-fit global
present-day star formation rates are $\rm 0.39 \, M_{\odot}/yr$ and $\rm 0.37 \, M_{\odot}/yr$, for the
LMC and SMC, respectively. Note that the smallest age bin used in the photometric analysis is 
$\rm log(t_{age}/yr) = 6.8$ ($6.6$) for the LMC (SMC) and that the uncertainties on the fit at
smaller stellar ages are quite large, with values in the range $\rm [0.25 - 0.82] \, M_{\odot}/yr$ 
($\rm [0.13 - 5.7] \, M_{\odot}/yr$) for the LMC (SMC).

Wide-field ground-based surveys, as MCPS, provide spatially comprehensive coverage, 
but the resulting CMDs only extend below the oldest main-sequence turnoff. As a 
result, the ancient ($> 4 - 5$~Gyr) SFH can not be well constrained. To overcome this limitation,
SFH studies have been done using the {\it Hubble Space Telescope} (HST), targetting - however -
only a limited number of fields (see e.g. Cignoni et al. 2012, 2013; Weisz et al. 2013). 
Comparison between different studies is made difficult by the use of different
assumptions regarding the stellar IMF, mass range, binary fraction and by the different
spatial coverage of the galaxies. In Cignoni et al. (2012), the authors have 
derived the SFH for the two most crowded regions of the SMC; in their Fig.~22, they show that 
the SFHs by Harris \& Zaritsky (2004)  for the same regions 
show a significantly higher stellar production prior to 8.4 Gyr ag1o. 
A comparison between the normalized cumulative star formation histories is shown in the left panel of 
Fig.\ref{fig:MCevo}; the solid lines with shaded region represent the predictions of Harris \& Zaritsky  and
the dashed lines are the recent results of Weisz et al. (2013) who used HST archival data. 
The LMC (SMC) formed 50 per cent of its mass $\sim 4 - 6$~Gyr ($\sim 3 - 4$~Gyr) ago and both
galaxies show a dramatic rise in SFR $\sim 3.5 - 4$~Gyr ago. Similar features are found in 
HST-based studies, although deeper data seem to suggested a larger cumulative star formation
history at earlier epochs. 

Using $\rm H_{\alpha}$ and $24 \mu$m emission to trace recent un-obscured and obscured star formation, 
Bolatto et al. (2011) have predicted a present-day global star formation rate of $\rm 0.037 M_{\odot}/yr$ for
the SMC. Following a similar approach, Skibba et al. (2012) have estimated present-day star formation rates
in the range $\rm [0.016 - 0.039] M_{\odot}/yr$ for the SMC and $\rm [0.25 - 0.63] M_{\odot}/yr$ for the LMC.
While the latter values are in good agreement with the prediction by Harris \& Zaritsky (2009), 
the former values seem to suggest a present-day star formation rate which is a factor of ten smaller than 
the results found by Harris \& Zaritsky (2004); even accounting for the large uncertainties associated to 
the best fit value, different SMC observations do not appear to be consistent. 

It is not surprising, therefore, that while the integrated star formation history of the LMC leads to a 
present-day stellar masses of $\rm M_{star} = 1.1 \times 10^9 M_{\odot}$, in good agreement with the 
value inferred from IR observations by Skibba et sl. (2012), the integrated stellar mass predicted
by Harris \& Zaritsky (2004) for the SMC is $\rm 1.21 \times 10^9 M_{\odot}$, almost a factor of 4 larger
than the value inferred from IR data (see the left panels of Figs.~\ref{fig:LMCmassevo} and \ref{fig:SMCmassevo}
in section 5). 

Hence, there are independent lines of evidence which seem to suggest that the recent star formation rate
predicted by Harris \& Zaritsky (2004) for the SMC may be over-estimated or that $\rm H_{\alpha}$ and 
IR observations may have missed a significant fraction of obscured star formation (although this seems
unlikely, given that the dust-to-gas ratio in the SMC is 1/1000, approximately a factor of 5 smaller
than the Milky Way, Leroy et al. 2007).    

In what follows, we will use the star formation histories for the LMC and SMC predicted by
Harris \& Zaritsky (2004, 2009), with the caveat that while the former appears to be
robust, the latter should be taken with caution, as different studies do not appear to converge
on the same result. 

In the center and right panels of Fig.~\ref{fig:MCevo} we compare the age metallicity relation 
implied by the metallicity-dependent star formation histories of Harris \& Zaritsky (2004, 2009) 
with the average data inferred by Carrera et al. (2008a, 2008b) using metallicities measured from 
spectroscopic data on individual stars (filled symbols) and with the results of Piatti (2012) and 
Piatti \& Geisler (2013) who used deep fotometric data on a large database of field stars (empty symbols).  
While the chemical evolution obtained by Harris \& Zaritsky (2004, 2009) is in good
agreement with metallicity and age measurements for star clusters, the best-fit age-metallicity relation
of the LMC appears to be lower than the observational data for field stars. The agreement improves
if the uncertainties due to the limited numbers of metallicity bins is considered. For this reason, in
what follows we will show the model predictions adopting the best fit star formation (and chemical 
evolution) histories and the corresponding lower and upper limits. 
     
\begin{figure*}
\includegraphics[width=80mm]{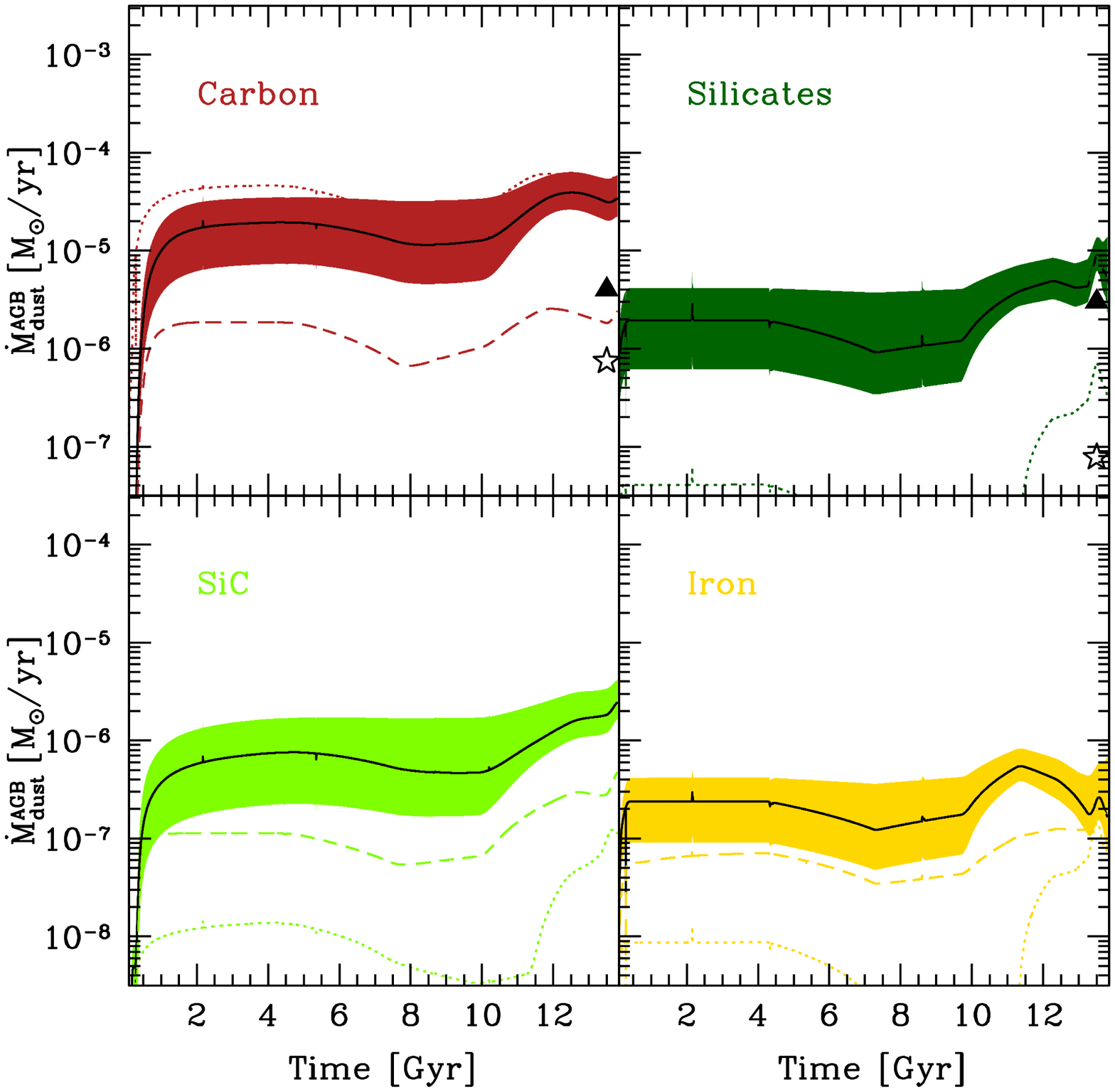}
\includegraphics[width=80mm]{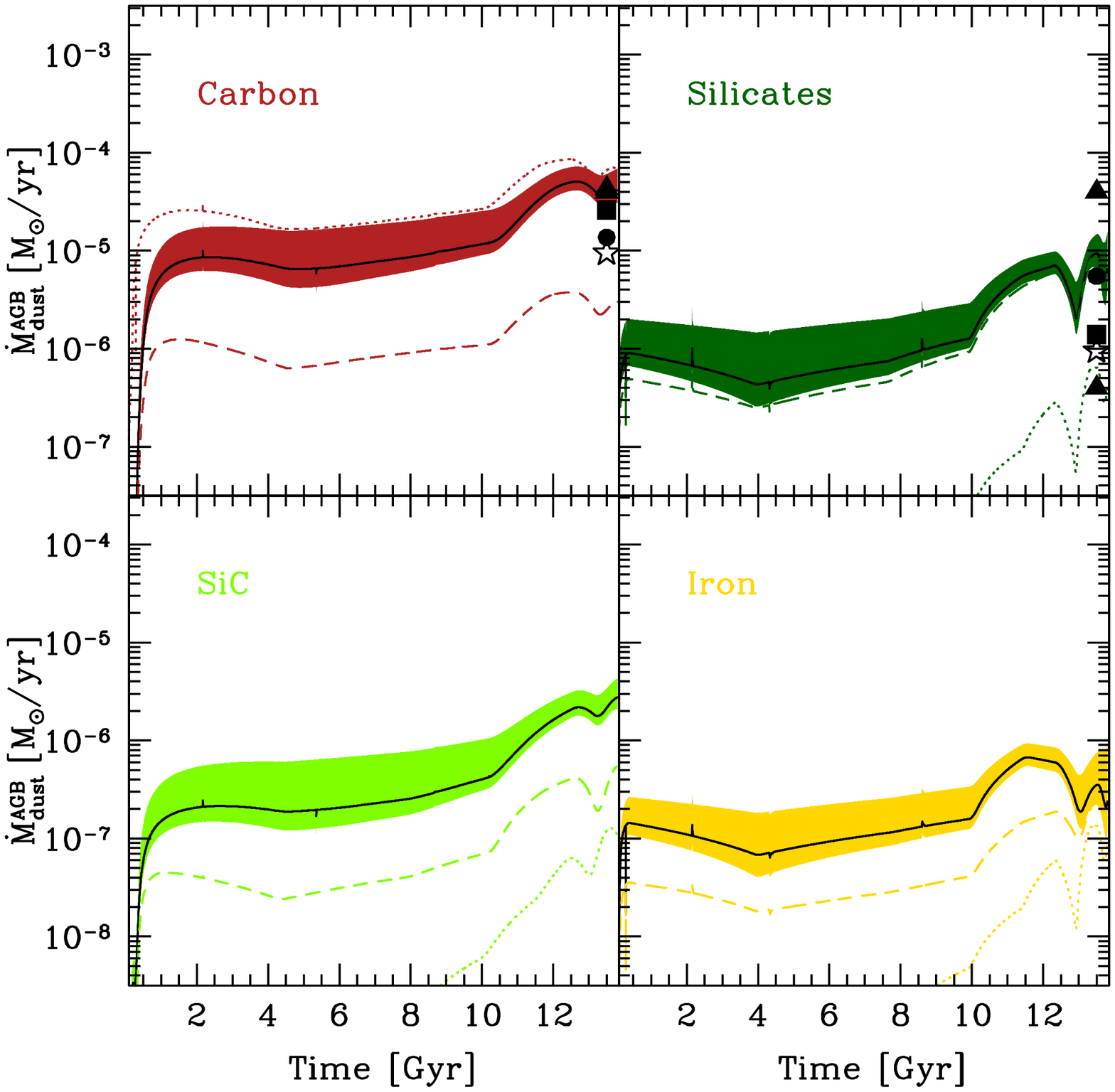}
\caption{The dust production rate (DPR) of AGB stars in the Small (left panel) and Large (right panel) Magellanic
Clouds as a function of time using the Harris \& Zaritsky (2004, 2009) metallicity-dependent star formation
histories. The solid lines show the predicted DPRs using the new ATON AGB yields, 
with shaded regions representing the uncertainty on the best-fit star formation histories. For comparison,
the dashed and dotted lines show the predicted DPRs for the best-fit star formation histories adopting
the old ATON yields and the Z08 yields, respectively. Data points show the DPRs
from C-AGB and O-AGB stars estimated by different studies and reported in Table~\ref{table:dpr}: Matsuura et al.
(2009, 2013, triangles), Srinivasan et al. (2009, squares), Boyer et al. (2012, stars), Riebel et al. (2012, circles).}
\label{fig:dpr}
\end{figure*}

\section{Dust production rate}
\label{sec:dprate}
Estimating the total dust production rate contributed by evolved stars requires the identification of 
point sources in a survey of a galaxy at near- and mid-infrared wavelengths, which are especially
suitable to study thermal dust emission. The Magellanic Clouds have been mapped by large photometric
surveys, such as the Magellanic Clouds Photometric Survey (MCPS, Zaritsky et al. 2004), the Two Micron
All Sky Survey (2MASS, Skrutskie et al. 2006), Surveying the Agents of a Galaxy's Evolution
Survey (SAGE) with the {\it Spitzer} Space Telescope (SAGE-LMC, Meixner et al. 2006; SAGE-SMC, Gordon et al. 2011), and
{\it HERschel} Inventory of The Agents of Galaxy Evolution (HERITAGE, Meixner et al. 2010, 2013).  
Spectroscopic follow-up to the SAGE surveys has provided fundamental information for the spectral classifications of the
point sources observed in SAGE (SAGE-Spec, Kemper et al. 2010). 

In principle, detailed radiative transfer modeling of each star should be carried out in order to 
accurately determine the rate of dust production around the star. In practice, this shows to be
very computationally expensive when samples of thousands of stars have to be analysed, 
although first applications on population studies have been attempted (Riebel et al. 2012). 
In general, however, measurements of the global dust input from large stellar samples 
rely on photometric techniques based on IR colours (Matsuura et al. 2009) or on IR excesses
(Srinivasan et al. 2009; Boyer et al. 2012).  

\begin{table*}
\begin{center}
\caption{Dust production rates by carbon and oxygen-rich AGB stars obtained in different analyses of the LMC and
SMC data. Where appropriate, sources classified as aO-AGBs have been included in O-AGBs and sources classified as 
x-AGBs have been included in C-AGBs. Note that in the recent analysis of Matsuura et al. (2013), the dust production
rates by O-AGBs is reported together with the contribution of Red SuperGiants (RSG), hence we have left these two
contributions together in the corresponding lines.}
\label{table:dpr}
\begin{tabular}{c|c|c|c}\hline
 & & Large Magellanic Cloud  &  \\ \hline
Sources  & $\rm \dot{M}_d [10^{-6} M_\odot/yr]$  & Number of Sources & Reference \\ \hline
C-AGB    & $43$ (up to 100)          		 & 1779              & Matsuura et al. (2009) \\
O-AGB    & $>> 0.4$ (expected $12$)  		 &           	     &       \\ \hline
C-AGB    & $26$                      		 & 7200      	     & Srinivasan et al. (2009) \\
O-AGB    & $1.4$		     		 & 8200      	     &                 \\ \hline
C-AGB    & $9.49$	          		 & 6076              & Boyer et al. (2012) \\
O-AGB    & $0.95$	   	  		 & 15243             &       \\ \hline
C-AGB    & $13.64 \pm 0.62$                      & 8049              & Riebel et al. (2012) \\
O-AGB    & $5.5 \pm 0.2$		         & 19566             &                 \\ \hline
C-AGB         & $40$	          		 & 906              & Matsuura et al. (2013) \\
O-AGB+RSG    & $40$	   	  		 & ...             &       \\ \hline
 & & Small Magellanic Cloud  &  \\ \hline
C-AGB         & $0.747$                      	& 1872              & Boyer et al. (2012) \\
O-AGB         & $0.078$		                & 3094              &                 \\ \hline
C-AGB         & $4$                      	& 399              & Matsuura et al. (2013) \\
O-AGB+RSG    & $3$		                & 86              &                 \\ \hline
\end{tabular}
\end{center}
\end{table*}
A comparison between the dust production rates (DPR) inferred by different studies is made
difficult by the use of different classification criteria, methods to infer the dust mass loss
rate, and dust opacitites. In most analyses, the following AGB stars sub-classes have been identified: 
carbon-rich (C-AGB) and oxygen-rich (O-AGB) AGB stars, 
{\it extreme} AGB stars (x-AGB), and anomalous O-rich (aO-AGB) sources 
(Cioni et al. 2006; Blum et al. 2006; Riebel et al. 2010; Boyer et al. 2011).
The latter are a sub-class of O-AGB stars while the x-AGB class is dominated by 
carbon stars but likely includes a small number of extreme O-rich sources. 
For the purpose of the present analysis, we have added together the aO-AGBs/O-AGBs and the C-AGBs/x-AGBs contributions 
and we show the corresponding dust production rate obtained by different analyses in Table \ref{table:dpr}. 

The theoretical DPR by AGB stars in the LMC and SMC can be computed using the star formation histories
for different metallicity bins described in section \ref{sec:sfr}. To be consistent with the analysis of 
Harris \& Zaritsky (2004, 2009), we adopt a Salpeter IMF in the stellar mass range $\rm 0.1 M_{\odot} \le m \le 100 M_{\odot}$. 

For a given grid of dust yields, $\rm m_{dust}(m,Z)$, we compute the time dependent total DPR as,
\begin{eqnarray}
\rm \dot{M}_{dust}(t) & = & \rm \Sigma_i \int_{m(t,Z_i)}^{m_{up}} dm' \, \Phi(m') \, m_{dust}(m', Z_i) \\ \nonumber
& & \rm SFR[t-\tau(m',Z_i),Z_i]
\label{eq:dpr}
\end{eqnarray} 
\noindent
where the index $\rm i$ runs over the metallicity bins, $\rm \tau(m,Z)$ is 
the mass and metallicity dependent stellar lifetime (Raiteri, Villata \& Navarro 1996), 
$\rm m_{up} = 100 M_\odot$, and the lower mass limit is such that $\rm \tau(m,Z_i) = t$. The contribution of AGB stars to the 
DPR, $\rm \dot{M}_{dust}^{AGB}$, is obtained setting the upper and lower integration limits 
to $\rm 0.1 M_{\odot} \le \rm m(t,Z_i) \le \rm 8 M_{\odot}$. We do not explicitly follow the metal enrichment of the galaxy; instead,
for each galaxy, we adopt the age metallicity relations implied by the metallicity dependent star formation histories. 

The predicted time evolution of the AGB DPR in the Magellanic Clouds is shown in Fig.~\ref{fig:dpr}, where we have separated the
contribution to carbon dust, silicates, SiC and iron grains production. The carbon dust and silicates 
production rates are compared with observational data of C-AGB and O-AGB stars (see Table~\ref{table:dpr}). 
For a given set of AGB stars dust yields, the time evolution of the DPRs is very similar in the two
galaxies, with present-day values for the best-fit star formation histories which differ by less than 30\%.

The three sets of AGB stars dust yields shown in the figures predict a similar evolution of the carbon DPR, although with
different amplitudes. Since AGB stars carbon dust yields do not depend significantly on the initial metallicity of the
progenitor stars, the features observed in the DPRs reflect analogous features in the global star formation history,
with a time delay whose amplitude depends on the lifetime of the most massive star producing carbon dust (about $7\,M_\odot$
for Z08 AGB stars yields and $3 \, M_\odot$ for ATON models). Both galaxies experienced a phase of
enrichment in the first 4 - 5 Gyr of evolution, then a quiescent epoch, followed by a second major episode
of enrichment roughly 2-3 Gyr ago and an enhanced DPR in the last few hundreds Myr of the evolution.   

Due to the increased mass loss rates in the C-star stage, the production rates of carbon grains predicted by
the new ATON yields is one order of magnitude larger than those predicted by the old ATON yields, 
and the resulting evolution is much closer to the predictions by Zhukovska \& Tielens (2013) based
on the Z08 AGB stars yields (see the dotted lines). However, while observations of C-AGB stars in the LMC seem to 
favour such larger mass loss rates, the opposite is true for the SMC where carbon dust production rates
predicted by the old ATON yields provides a better match to the observational data. 

The production rates of silicates predicted by the ATON yields (note that the old and new yields are
consistent within the uncertainties due to the star formation histories) are in good agreement with the observations
of O-AGB stars both in the LMC and in the SMC. Similarly good agreement is found when Z08 AGB stars yields are adopted.
However, in this case silicate dust production is effective only in the last 2 Gyrs of the evolution. 
Conversely, efficient HBB in ATON models allows silicate dust production since the earliest evolutionary
times, as stars with $\rm m_{star} > 3 \, M_\odot$ form silicate grains already at a metallicity $\rm Z = 0.001$.

The production of iron grains follows a time evolution similar to that of silicate grains, as these
two grain species form under HBB conditions in ATON models, although with different rates. Conversely, 
the enrichment in SiC closely follows the evolution of carbon DPR, but with a smaller amplitude, reflecting
the relative abundance of these two species in the dust yields produced by low-mass stars at any metallicity.

\begin{figure*}
\includegraphics[width=80mm]{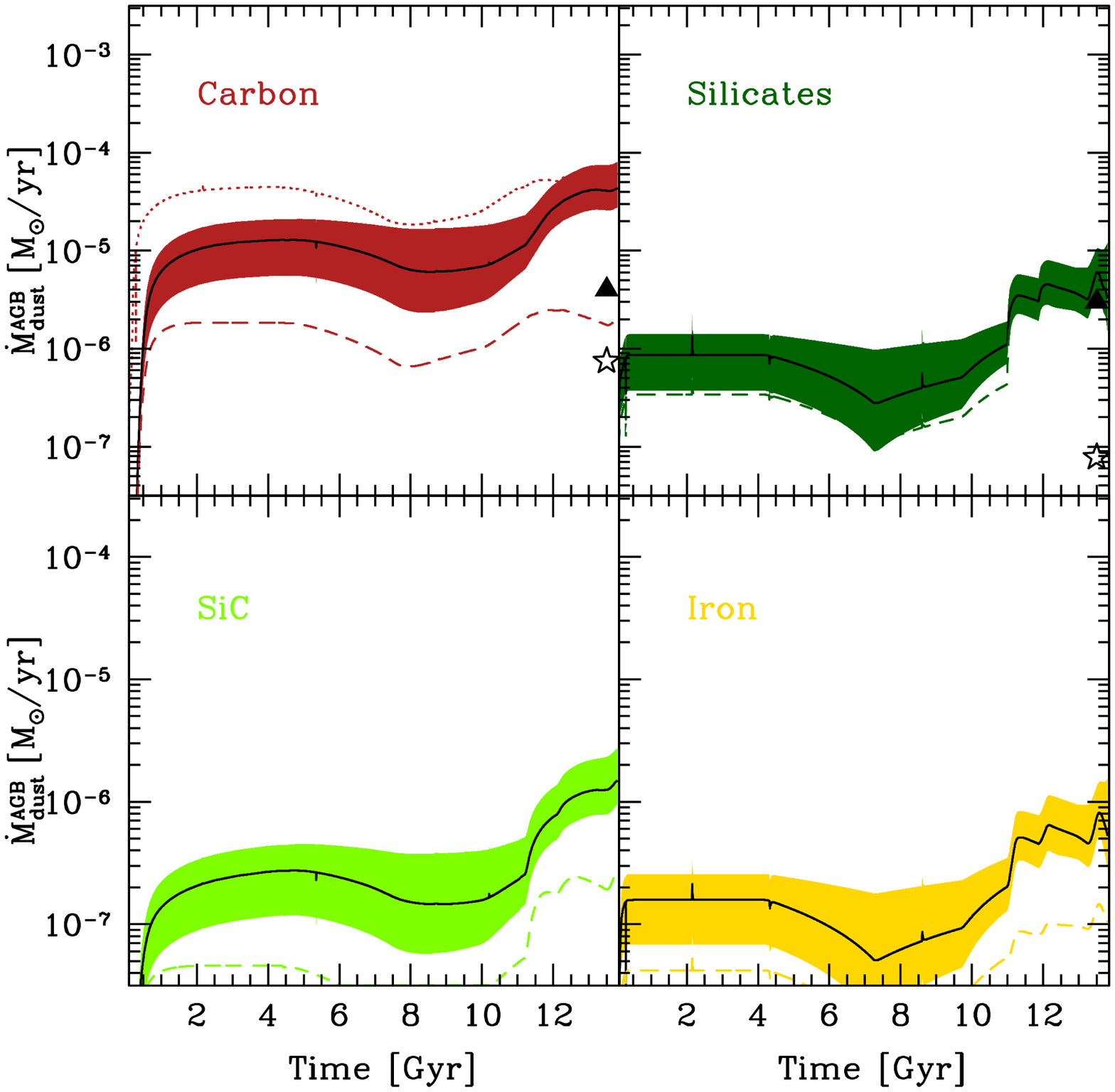}
\includegraphics[width=80mm]{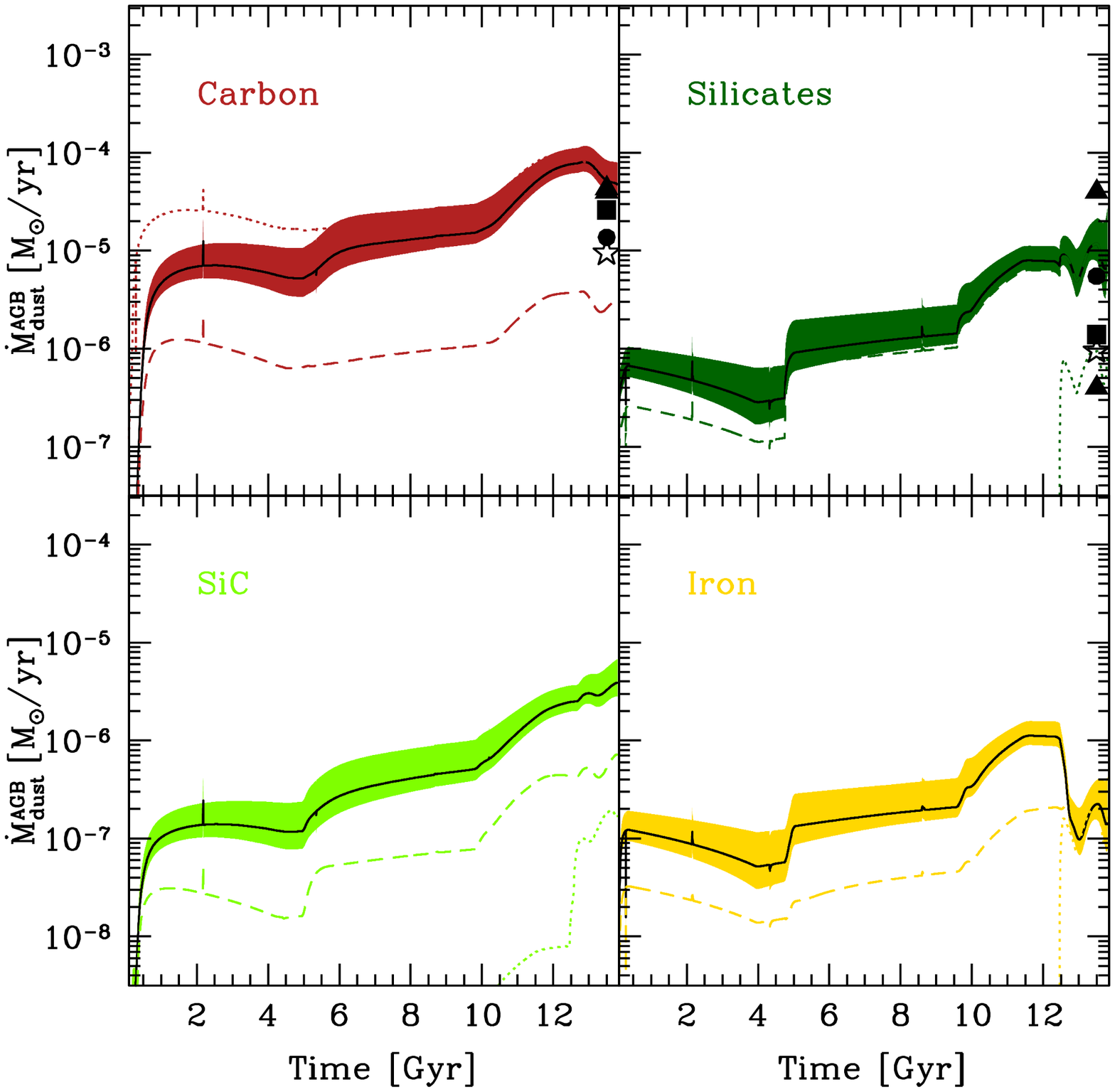}
\caption{Same as in Fig.\ref{fig:dpr} but adopting 
the Harris \& Zaritsky (2004, 2009) global star formation
histories and the Piatti (2012) and Piatti \& Geisler (2013) age metallicity
relations for the SMC and LMC, respectively, to estimate the metal enrichment
history of the ISM. In the left panel, silicates, SiC and Iron DPRs for the 
Z08 AGB stars yields (dotted lines in Fig.~\ref{fig:dpr}) are much below
the plotted range (see text).}
\label{fig:dprpiatti}
\end{figure*}

Note that with the adopted star formation histories, the carbon DPR in the last 4 Gyr of the evolution is
dominated by $Z = 0.004$ AGB stars and that in both galaxies $Z=0.008$ stars dominate only in the last 
few hundreds Myr of the DPR evolution. Given the small sensitivity of theoretical carbon dust
yields to the metallicity of the progenitors, we do not expect a strong dependence on the 
adopted metal enrichment history for the galaxies. 

We test this hypothesis by adopting a different metallicity evolution: we use 
the age metallicity relations inferred by Piatti \& Geisler (2013) for the LMC and 
by Piatti (2012) for the SMC (see Fig.~\ref{fig:MCevo}) as a proxy for the metal enrichment of the ISM
of the galaxies; hence, we can recompute the DPRs implied by the total star formation histories of 
Harris \& Zaritsky, without differentiating among the different metallicity-dependent components.
As expected, Fig~\ref{fig:dprpiatti} shows that the carbon DPRs predicted by a given set of 
AGB stars dust yields do not differ significantly from the ones shown in Fig.~\ref{fig:dpr}. 

Conversely, the lower metallicities implied by the age metallicity relations 
(particularly for the SMC, where there is no single stellar population with metallicity $\rm Z > 0.004$) 
result in a significant depression of the silicate DPRs predicted by Z08 AGB stars yields, 
which are below the observational data for the SMC by several orders of magnitudes.

Hence, we conclude that current samples of AGB stars in the LMC favour strong mass loss
rates from C-stars, as predicted by the new ATON yields with the Wachter et al. (2008)
mass loss rate prescription, and by the Z08 AGB stars yields. 
The same yields, however, exceed the DPRs observed for C-AGB stars in the SMC by approximately
one order of magnitude. The latter data are better reproduced by the old ATON yields, which were
based on the less efficient mass loss rate prescription by Bl\"ocker et al. (1995). 

This may be an indication of a metallicity dependence of the mass loss rates during the
carbon-star phase of AGB evolution. We note that Bl\"ocker et al. (1995) based their prescription
on a description of the circumstellar envelope of MIRA variables that neglects 
the effects of radiation pressure on dust, probably underestimating mass loss rates in the 
carbon-stage phase. Conversely, the formulation by Wachter et al. (2008) 
is based on hydrodynamical wind models which include carbon-dust formation; 
however, there is no significant metallicity dependence in their formulation
and the resulting mass-loss rates are of the same order of magnitudes for solar metallicity
models, models with the metallicity of the LMC, and models with the SMC metallicity.

A unique feature of the ATON yields is that efficient HBB experienced by stars with $\rm m > 3 M_{\odot}$
lead to silicate dust production by AGB stars with initial metallicity $\rm Z \ge 0.001$. Conversely,
both Zhukovska et al. (2008) and Nanni et al. (2013) find that silicate grains can only form in
AGB stars when their initial metallicities are higher; for the metallicity range relevant to the
present analysis, silicate yields from Z08 AGB stars are not negligible only for stellar metallicity $\rm Z = 0.008$.
Hence, if the dominant stellar population in the SMC have metallicities $\rm Z \le 0.004$, as observed by
Piatti \& Geisler (2013), the observed population of O-AGB stars suggests that silicate dust formation
should occurr already at lower metallicities, as predicted by efficient HBB occurring in the ATON models.  

Finally, since the total DPRs are dominated by carbon grains, their current values are not significantly
affected by the metal enrichment history and depend more on the adopted dust yields. 
The total DPR by AGB stars in the LMC is predicted to be $\rm [9.86, 51.3, 67.1] \times 10^{-6} \, M_\odot/yr$ for the 
ATON, new ATON, and the Z08 AGB stars dust yields. Although these values span the range 
indicated by observations of C-AGB and O-AGB stars, the new ATON and the Z08 yields lead to DPRs which are larger
than most of the observational result (except for the most recent Matsuura et al. 2013 determinations, which 
however refer to the global contributions of C-AGB, O-AGB stars and RSGs). 
Similarly, the total DPR by AGB stars in the SMC is predicted to be $\rm [7.62, 41.7, 48] \times 10^{-6} \, M_{\odot}/yr$
for the ATON, new ATON, and the Z08 AGB stars dust yields, with the latter two dust yields exceeding the observed DPRs by one
order of magnitude. 

How do these results depend on the uncertainties in the SMC star formation rate discussed in section~\ref{sec:sfr}?
Even if we were to scale-down\footnote{ 
The scale factor was chosen so as to reproduce the present-day global stellar mass in the SMC 
and - at the same time - have a recent star formation rate in agreement with $\rm H_{\alpha}$ and $24 \mu$m observations 
(at least within the uncertainties).} the star formation rate predicted by Harris \& Zaritsky (2009) by a factor 4, the present-day carbon DPR predicted by the new ATON AGB stars yields (and by Z08 AGB stars yields) 
would be $\rm \approx 10^{-5}\,M_\odot/yr$,  larger than the observational data on C-AGB stars by 
Boyer et al. (2012) and Matsuura et al. (2013). The old ATON yields would decrease the DPR 
to $\rm 5.43 \times 10^{-7} \, M_{\odot}/yr$, still in agreement with observations. 

In the following, we will estimate the contribution of AGB stars to the total stellar
dust budget if the MCs using the new ATON and the old ATON models as the reference dust yields for the LMC
and SMC, respectively.

\begin{figure}
\includegraphics[width=80mm]{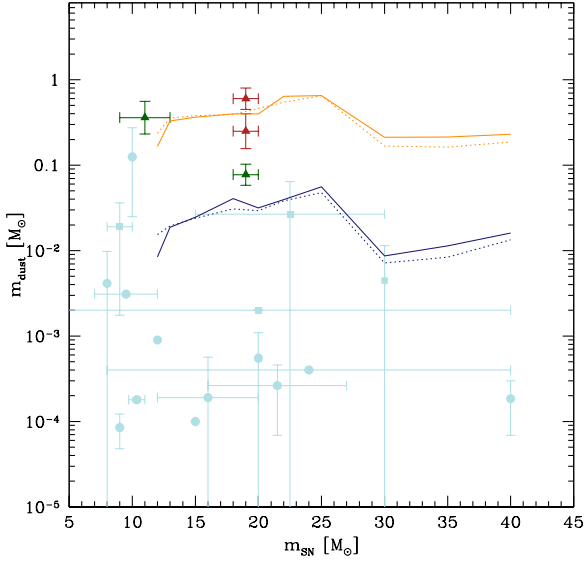}
\caption{Mass of dust produced by supernovae as a function of the progenitor mass.
The data points in light blue are taken from the compilation published in Gall, Hjorth \&
Andersen (2011) and Otsuka et al. (2012): circles refer to observations of SNe (12 
objects) and squares to observations of SN remnants (4 objects). 
The triangles indicate Herschel detected sources in the Milky Way (dark green) and in the LMC (red, see text).
The two sets of lines represent dust
yields predicted by theoretical models for stars with initial metallicities $\rm Z = Z_{\odot}$
(solid) and $\rm Z = 0.1 Z_{\odot}$ (dotted); the upper (lower) set of lines shows the mass of dust 
before (after) the passage of the reverse shock.}
\label{fig:snmass}
\end{figure}

\section{Total dust budget}
\begin{figure*}
\includegraphics[width=58mm]{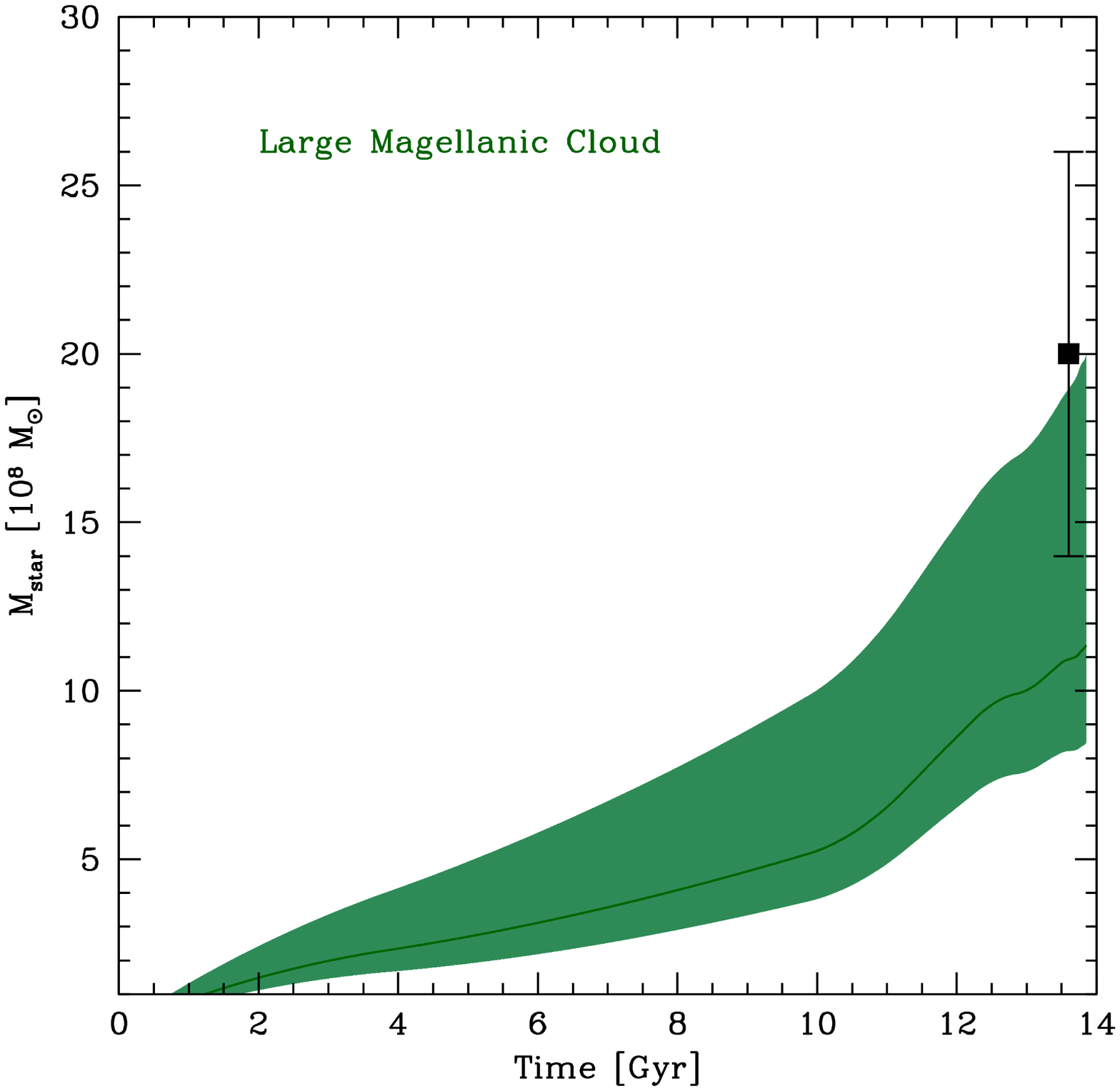}
\includegraphics[width=58mm]{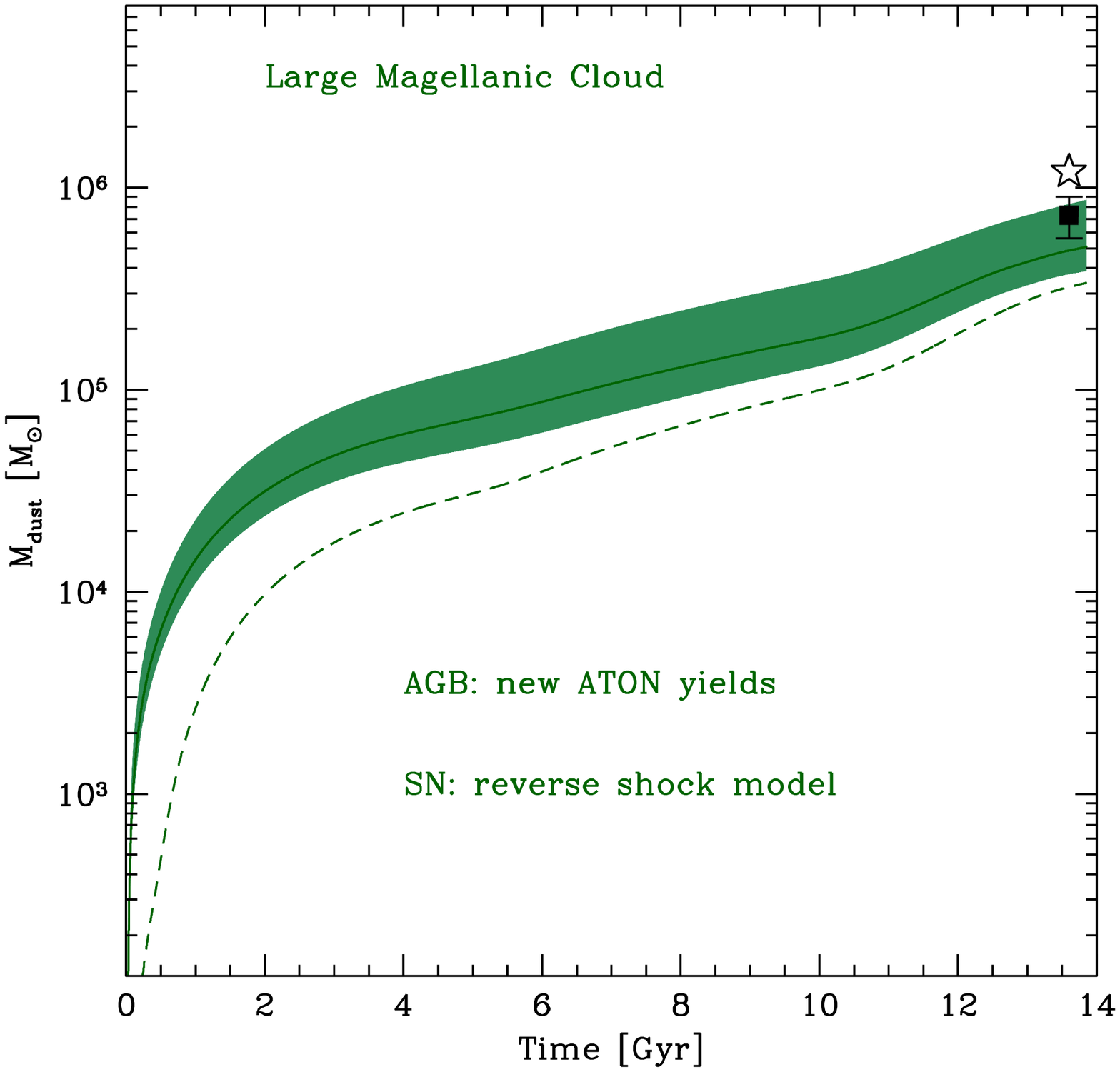}
\includegraphics[width=58mm]{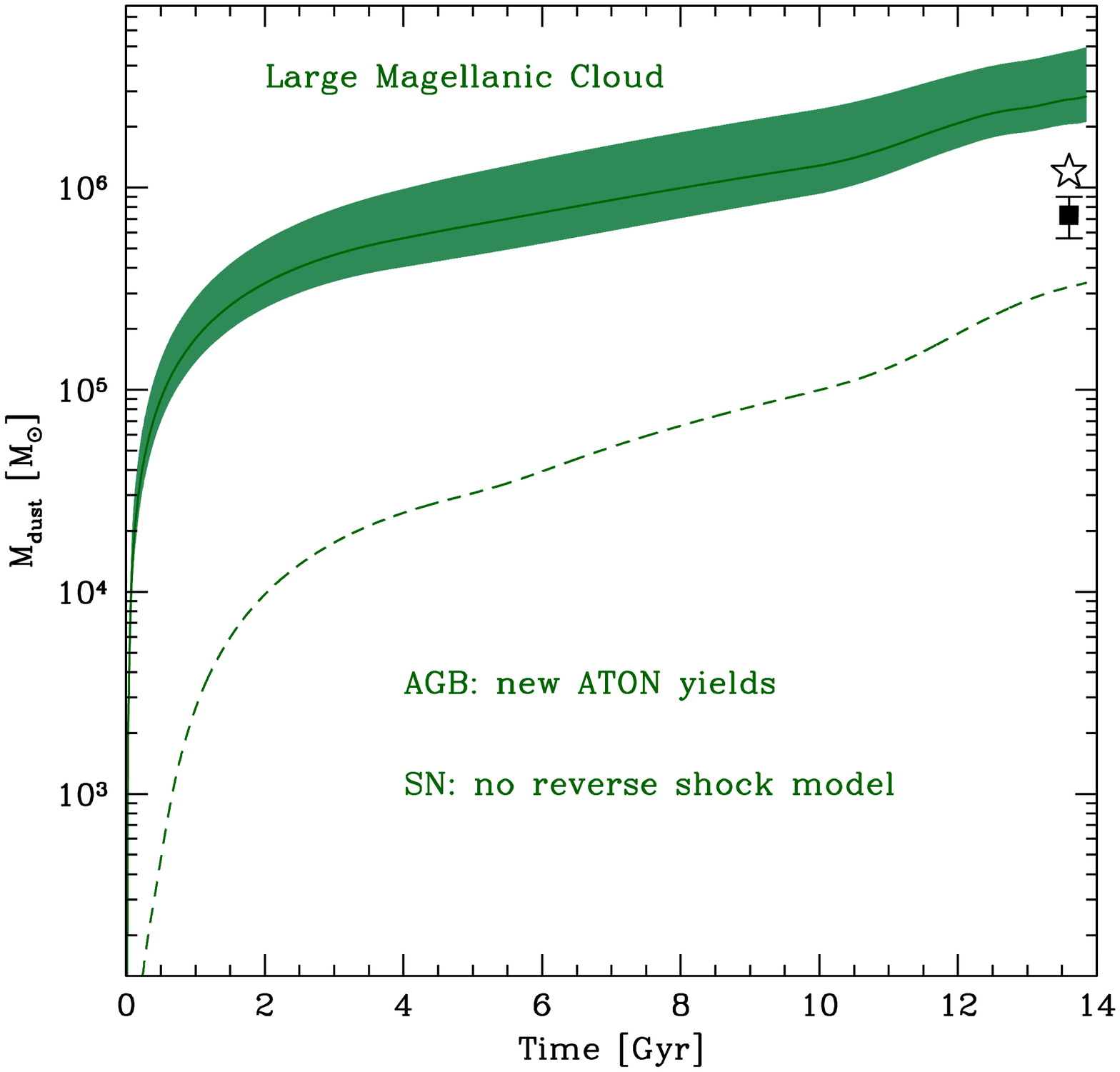}
\caption{Time evolution of the stellar (left panel) and dust masses (central panel) predicted by the LMC model.
We used the Harris \& Zaritsky (2009) global star formation histories (solid lines indicate the predictions 
obtained adopting the best-fit SFH and the shaded regions illustrate the uncertainty on the fit) 
and the Piatti \& Geisler (2013) age metallicity relation to estimate the metal enrichment
history of the ISM. The observational data are taken from Skibba et al. (2012, star) and from Gordon et al. (2014, square). 
The right panel shows the dust mass evolution adopting the no reverse shock models for the dust yields of 
massive stars (see Fig.\ref{fig:snmass}). The dashed lines show the mass of dust produced by 
AGB stars only, adopting the best-fit SFH.}
\label{fig:LMCmassevo}
\end{figure*}

\begin{figure*}
\includegraphics[width=58mm]{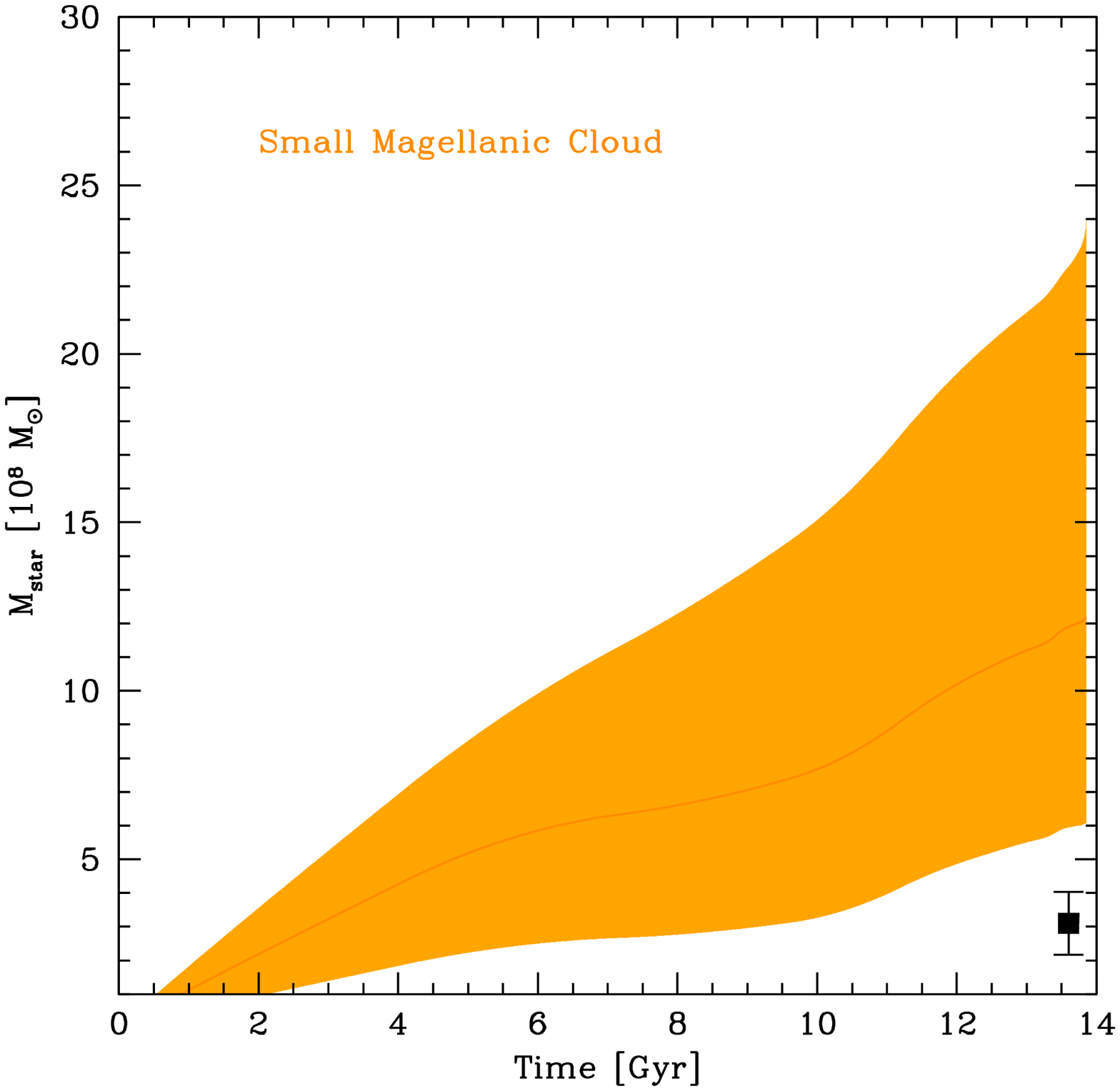}
\includegraphics[width=58mm]{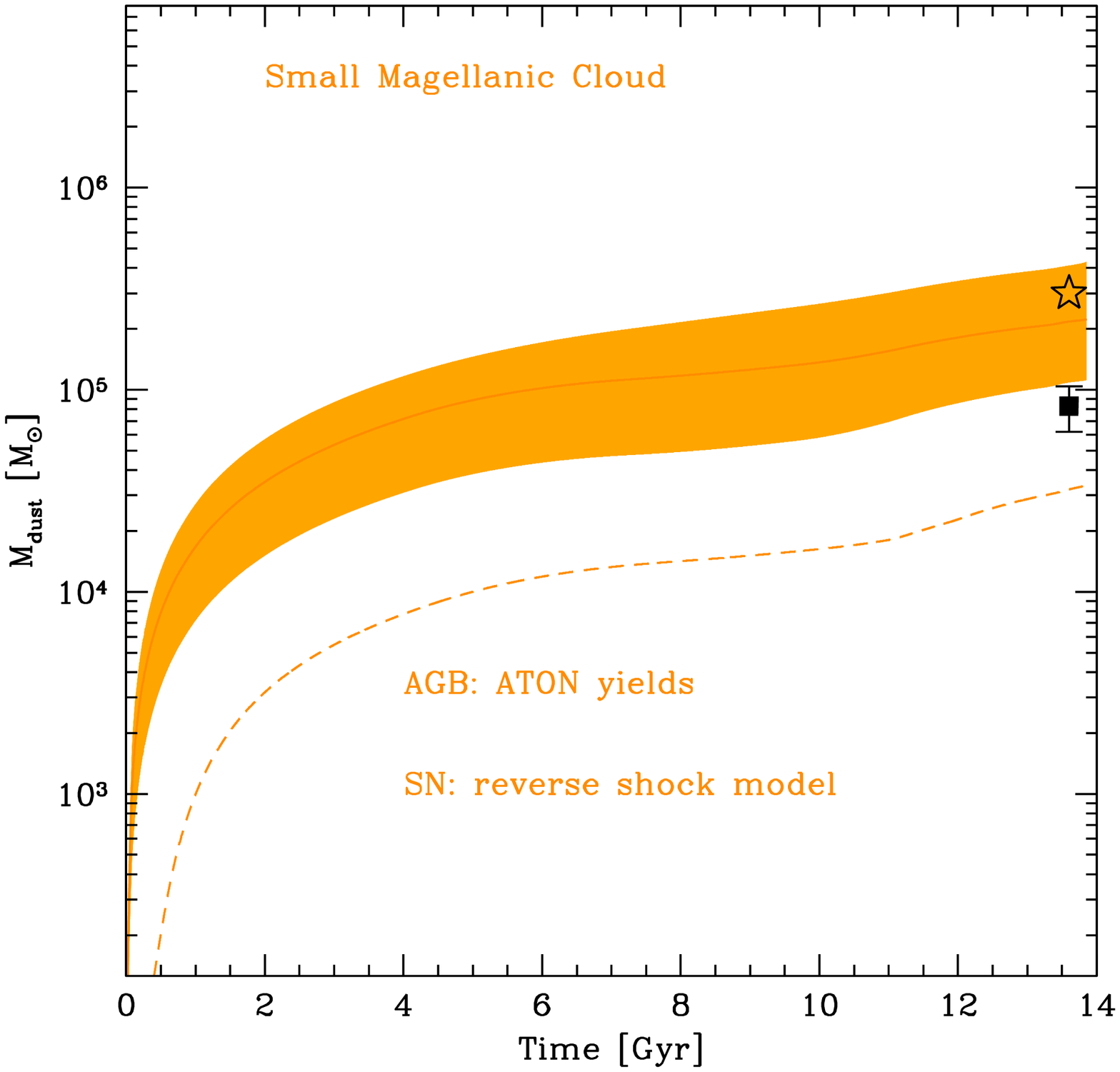}
\includegraphics[width=58mm]{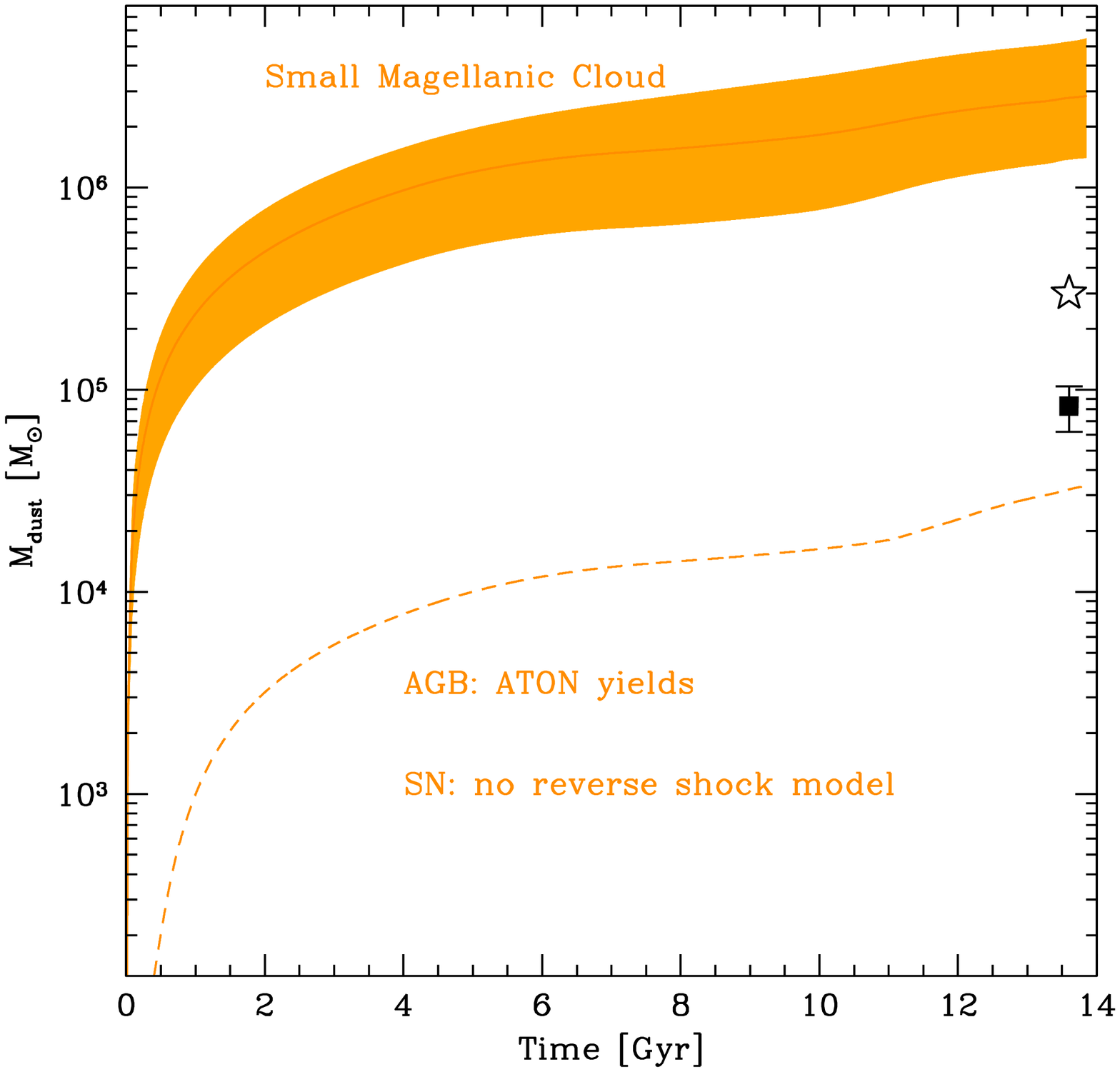}
\caption{Same as Fig.\ref{fig:LMCmassevo} but for the SMC model.
We used the Harris \& Zaritsky (2004) global star formation histories (solid lines indicate the predictions 
obtained adopting the best-fit SFH and the shaded regions illustrate the uncertainty on the fit) 
and the Piatti (2012) age metallicity relation to estimate the metal enrichment
history of the ISM.} 
\label{fig:SMCmassevo}
\end{figure*}

The total mass of dust present in the interstellar medium of the LMC and SMC has
been estimated using different techniques (extinction, emission, elemental depletions)
and observational facilities (see Table~1 in Meixner et al. 2013 for a recent compilation). 
Total dust masses of $\rm M_{dust} = 1.2 \times 10^{6} M_{\odot}$ and $3\times 10^5 M_{\odot}$ 
have been derived through analyses of {\it Spitzer} SAGE maps of the LMC and SMC, respectively 
(Bernard et al. 2008; Leroy et al. 2007). Recently, Skibba et al. (2012) have provided new
estimates of the total dust mass using {\it Herschel} HERITAGE data; after convolving the 
photometric data for the two galaxies to a common resolution, the authors inferred a total
dust mass of $\rm M_{dust} = 1.1 \times 10^{6} M_{\odot}$ for the LMC 
and $1.1\times 10^5 M_{\odot}$ for the SMC, in good agreement with previous estimates.
 At the time of paper submission, a new analysis of the dust surface density maps
of the MCs by the HERITAGE team was completed (Gordon et al. 2014) and the
resulting integrated dust masses were $\rm (7.4 \pm 1.7)\times 10^5 M_{\odot}$ and 
$\rm (8.3 \pm 2.1)\times 10^4 M_{\odot}$ for the LMC and SMC, respectively.
These values of the dust masses are significantly smaller than previous estimates,
probably due to different assumptions in the models used to fit the maps.


In the following analysis, we will compare {\it Herschel}  data 
to the mass of dust produced by AGB stars and supernovae (SN) adopting the star formation 
and metal enrichment histories discussed in the previous sections. 
We take the new ATON (old ATON) yields to compute the mass of dust 
produced by AGB stars in the LMC (SMC) as these provide a good match
to the observed AGB dust production rates. 

Dust yields for massive stars have been theoretically calculated by different groups 
(Todini \& Ferrara 2001; Nozawa et al. 2003; Bianchi \& Schneider 2007; Cherchneff \& Dwek 2010;
Sarangi \& Cherchneff 2012). The effective SN dust yields, i.e. the mass
of dust that is able to survive the passage of the reverse shock enriching
the ISM, can be significantly smaller than the dust mass newly formed in the
expanding ejecta (Bianchi \& Schneider 2007; Nozawa et al. 2007; Silvia et al. 2010). 
In Fig.\ref{fig:snmass} we compare theoretical models with observational data. 
 Dust masses of the same source obtained by different groups have been 
averaged and represented by a single data point.
Herschel-detected sources are marked as triangles: Crab and CasA,
(dark green, Gomez et al. 2012; Dunne et al. 2009; Barlow et al. 2010) and  SN1987A 
and N49 (red, Matsuura et al. 2011; Otsuka et al. 2010). The latter data has to be viewed as an
upper limit to the newly formed dust due to possible contamination from ISM dust.
We consider Herschel-detected sources to provide a more complete census of the 
dust associated to the SN, being sensitive to the dynamically dominant cool dust 
component. 
Yet, none of the remnants is old enough (ages $< 10^3$ yr) for the reverse shock
to have significantly affected the newly formed dust. Given this uncertainty,
we will compute the mass of dust produced by massive stars adopting the 
Bianchi \& Schneider (2007) dust yields for the no-reverse 
shock case shown with the upper lines and with a reverse shock model 
where the explosions take place in circumstellar
medium with average density of $\rm n_{ISM} = 1 cm^{-3}$ (lower lines).

In what follows, our aim is to compare the maximum contribution to the existing dust mass  by stellar sources.
Hence, we simply integrate Eq.~1, without considering the effective dust lifetime in the ISM, due to
the combined effects of destruction by interstellar shocks in the hot diffuse ISM or the process of grain growth by accretion of 
gas-phase elements in the dense ISM (see e.g. Valiante et al. 2011, 2014 for a complete chemical evolution model
with dust). The integrated dust masses that we compute here are to be viewed as the maximum dust budget that
stellar sources may potentially contribute.

The results for the LMC  are shown in Fig.\ref{fig:LMCmassevo}. 
In the left panel, we plot the time evolution of the 
stellar mass\footnote{The stellar mass shown in the figure is not simply the time integral of the star formation
rate, but takes into account the fraction of stellar remnants prediced by the adopted Salpeter IMF. Hence, it can
be viewed as the mass of active stars to be compared to the stellar mass inferred by the spectral energy distribution.}
predicted by the Harris \& Zaritsky (2009) star formation rate. The observational data points are taken from Skibba et al. (2012),
who used the calibration of Eskew et al. (2012) to convert from the 3.6 and 4.5 $\mu$m flux densities 
to stellar mass, finding  ${\rm M_{star}} = 2 \times 10^9 \rm M_{\odot}$ with approximately $30\%$ uncertainty.
It is clear that the predicted stellar mass is consistent with the observational data point but only adopting the
upper limit on the SFH from the Harris \& Zaritsky (2009) analysis. 

The total mass of dust produced by stellar sources is too small to account for the present-day
dust mass in the LMC derived by Skibba et al. (2012) but it is in good agreement with the more recent
estimates by the HERITAGE team (Gordon et al. 2014),  if no significant destruction in the ISM occurs. This is shown in the central panel
of the same figure, where we compare the mass of dust predicted by the model adopting 
the new ATON yields for AGB stars and the reverse shock yields for SNe. The total mass of dust produced by stellar
sources ranges between $\rm 3.9 \times 10^5 M_{\odot}$ and $\rm 8.7 \times 10^5 M_{\odot}$, with the best-fit 
value of $\rm M_{dust} =  5.1 \times 10^5 M_{\odot}$.
While the predicted dust masses are consistent with previous findings by Matsuura et al. (2009) and Zhukovska \& Henning (2013),
the previously reported large discrepancies between dust input from stars and the existing interstellar dust mass are no longer
supported by the most recent data.
In the conservative limit of reverse shock yields, massive stars do not represent the dominant dust sources, with 
AGB stars contributing to $67\%$ of the final dust mass.

Similar conclusions apply for the SMC, although with larger uncertainties (see Fig.~\ref{fig:SMCmassevo}). In fact, 
the star formation history by Harris \& Zaritsky (2004) predicts a stellar mass 
larger than the value inferred by Skibba et al. (2012), $\rm M_{star} = 3.1 \times 10^8 M_{\odot}$.
Although the latter value has been derived using a calibration based on a detailed analysis of
the LMC (Eskew et al. 2012), the difference between the final stellar masses is larger than
the observational uncertainty. With this caveat, stellar sources in the SMC model produce 
a total mass of dust in the range $1.1 \times 10^5 \rm M_{\odot} \leq \rm M_{dust} \le \rm 4.3 \times 10^5 M_{\odot}$,
with a best-fit value of $\rm M_{dust} = 2.3 \times 10^5 M_{\odot}$. This is in good agreement with the observed
mass of dust in the SMC, with massive stars dominating the total dust budget at all times and AGB stars 
producing only $15\%$ of the total ISM dust mass. Note that
if we were to scale down the best-fit star formation rate of Harris \& Zaritsky (2009) in order
to reproduce the observed stellar mass (see section~\ref{sec:dprate}), the total dust mass produced by stars would be
in $\rm 2.8 \times 10^4 M_{\odot} \le \rm M_{dust} \le \rm 10^5 M_{\odot}$, with a best-fit value of $5.6 \times 10^4 M_{\odot}$.
These values appear to be more consistent with the results by 
Boyer et al. (2012) and Matsuura et al. (2013), who estimate the lifetime of dust in the SMC 
using star formation rates in the range $[\rm 0.037 - 0.08 M_{\odot}/yr]$, 
lower than the results by Harris \& Zaritsky (2009) and closer to the values reported 
by Bolatto et al. (2011) and Skibba et al. (2012).  Yet, while previous studies concluded that 
the dust input from stars in the SMC was too low to account for the existing interstellar dust mass, we find that
the most recent observations by the HERITAGE team (Gordon et al. 2014) support a stellar origin for dust in the SMC.

Finally, we have recomputed the dust mass evolution assuming that dust production in SN does not suffer
any destruction by the reverse shock (no reverse shock models). The right panels of Figs.~\ref{fig:SMCmassevo} 
and \ref{fig:LMCmassevo} show that the mass of dust produced by stars {\it exceeds} the observational data in both
galaxies. It is important to stress that all the above findings have been obtained assuming that 
dust grains injected by stars in the ISM do not suffer further reprocessing, like destruction by
interstellar shocks or grain growth in dense clouds. Hence, our analysis suggests that moderate
destruction by the reverse shock of the SNe, or by interstellar shocks might have occurred during
the evolution of the galaxies. A detailed investigation of dust reprocessing in the ISM requires
the development of a full chemical evolution model with dust (Valiante et al. 2009, 2011, 2014),
including a two-phase description of the dense and diffuse phases of the ISM (de Bennassuti et al. 2014),
which goes beyond the scope of the present analysis.    

\section{Conclusions}

In this paper, we have compared theoretical dust yields for AGB stars to observations of dust production
rates by carbon-rich and oxygen-rich AGB stars in the Small and Large Magellanic Clouds. Our aim is to test
whether current observations have the potential to discriminate among different models and to shed light
on the complex dependence of the dust yields on the mass and metallicity of progenitor stars.

Using metallicity dependent star formation histories inferred by Harris \& Zaritsky (2004, 2009) based
on the MCPS survey, we find that:
\begin{itemize}
\item Observed dust production rates by carbon-rich AGB stars in the LMC favour theoretical models
with strong mass loss, as predicted by the new ATON models with Wachter et al. (2008)
mass loss prescription.
\item The same yields, however, exceed the dust production rate observed for carbon-rich AGB stars
in the SMC by approximately one order of magnitudes. Hence, current data of the SMC seem to favour
the old ATON yields, which were based on the less efficient mass loss rate prescription by 
Bl{\"o}cker et al. (1995). This conclusion is independent of the uncertainties associated to the
star formation or the metal enrichment histories of the galaxy and may be an indication of a 
stronger metallicity dependence of the mass-loss rates during the carbon-star stage.
\item Efficient Hot Bottom Burning in ATON models allows stars with $\rm m_{stars} > 3 \, M_{\odot}$ to 
enrich the interstellar medium with silicate and iron grains already at very low metallicities, $\rm Z \ge 0.001$,
at odds with dust yields based on synthetic AGB models. If the dominant stellar populations in the SMC
have metallicities $\rm Z \le 0.004$, observations of dust production rates by oxygen-rich stars have the
potential to confirm or refute the theoretical predictions.
\item The latest analysis by the HERITAGE team (Gordon et al. 2014) leads to integrated dust masses in the LMC and
SMC that are a factor 2-4 smaller than previous estimates. When compared to our model predictions,
we find that the existing dust mass in the ISM of the MCs can have a stellar origin, even without
resorting to extreme yields, unless significant
destruction of the newly formed dust in SN reverse shock or in the ISM takes place.
\end{itemize}

Our study confirms the potential of the Magellanic Clouds as fundamental astrophysical laboratories
to test our current understanding of the dust cycle in the interstellar medium. Yet, conclusions
based on detailed comparison between models and observations are hampered by uncertainties on the
star formation and chemical enrichment history of the galaxies, particularly of the Small Magellanic
Cloud. Future observational studies complemented by a more realistic modelling of the two brightest
satellites of the Milky Way in a cosmological context (Boylan-Kolchin et al. 2011) are needed to 
make substantial progress.

\section*{Acknowledgments}
We thank  Alberto Bolatto, Michele Cignoni and Mikako Matsuura for their kind collaboration and for useful discussions.
The research leading to these results has received funding from the European Research Council under the European Union's 
Seventh Framework Programme (FP/2007-2013) / ERC Grant Agreement n. 306476.
HH thanks support from the NSC grant NSC102-2119-M-001-006-MY3. FK acknowledges financial support from the NSC 
under grant number NSC100-2112-M-001-023-MY3.

\label{lastpage}

\end{document}